\documentclass[twocolumn]{aastex62}

\graphicspath{{./}{figures/}}

\received{}
\revised{}
\accepted{}
\submitjournal{ApJ}

\shorttitle{}
\shortauthors{R. Mart\'in-Dom\'enech et al.}
\begin{document}

\title{%Formation of COMs in interstellar ice mantles. 

Formation of NH$_2$CHO and CH$_3$CHO upon UV photoprocessing of interstellar ice analogs}

\correspondingauthor{Rafael Mart\'in-Dom\'enech}
\email{rafael.martin\_domenech@cfa.harvard.edu}

\author{Rafael Mart\'in-Dom\'enech}
\affil{Center for Astrophysics $|$ Harvard \& Smithsonian \\
60 Garden St., Cambridge, MA 02138, USA}

\author{Karin I. \"Oberg}
\affil{Center for Astrophysics $|$ Harvard \& Smithsonian \\
60 Garden St., Cambridge, MA 02138, USA}

\author{Mahesh Rajappan}
\affil{Center for Astrophysics $|$ Harvard \& Smithsonian \\
60 Garden St., Cambridge, MA 02138, USA}

\begin{abstract}

Complex organic molecules (COMs) can be produced by energetic processing of interstellar ice mantles accreted on top of dust grains. 
Two COMs with proposed energetic ice formation pathways are formamide and acetaldehyde. 
Both have been detected in Solar System comets, 
and in different circumstellar and interstellar environments. 
In this work, we study the NH$_2$CHO and CH$_3$CHO formation upon UV photoprocessing of CO:NH$_3$ and CO:CH$_4$ ice samples. 
The conversion from NH$_2^.$ radicals to NH$_2$CHO is 2$-$16 times higher than the conversion from CH$_3^.$ radicals to CH$_3$CHO under the explored experimental conditions, likely because the formation of the latter competes with the formation of larger hydrocarbons.  
In addition, the conversion of NH$_2^.$ into NH$_2$CHO at 10 K increases with the NH$_3$ abundance in the ice, and also with the temperature in CO-dominated CO:NH$_3$ ices. 
%IM NOT ONLY REFERRING TO THE EXPERIMENTS PERFORMED AT 20K, BUT ALSO TO THE INCREASED CONVERSION UPON WARM-UP OF THE ICES IRRADIATED AT 10K.
This is consistent with the presence of a small NH$_2^.$ and HCO$^.$ reorientation barrier 
%Both trends suggest that a particular orientation of the NH$_2^.$ and HCO$^.$ radicals is needed 
for the formation of NH$_2$CHO, 
%FOR SOME REASON, THIS ORIENTATION IS MORE EASILY ACCESED WHEN NH3 IS MORE ABUNDANT. 
%and that there is a small reorientation barrier.  
%THE REORIENTATION BARRIER CAN BE SAVED AT SLIGHTLY HIGHER TEMPERATURES. 
which is overcome with an increase in the ice temperature. 
%No evidences of a similar barrier were found for the formation of CH$_3$CHO. 
%
The measured NH$_2$CHO and CH$_3$CHO formation efficiencies and rates are similar to those found during electron irradiation of the same ice samples under comparable conditions, %(COMPARABLE MEANING SIMILAR DEPOSITED ENERGY)
suggesting that both UV photons and cosmic rays would have similar contributions to the solid-state formation of these species in space. 
Finally, the measured conversion yields (up to one order of magnitude higher for NH$_2$CHO) suggest that in circumstellar environments, where the observed NH$_2$CHO/CH$_3$CHO abundance ratio is $\sim$0.1,  there are likely additional ice and/or gas phase formation pathways for CH$_3$CHO. 
%CH$_3$CHO production must be dominated by other pathways. 

\end{abstract}



\section{Introduction} \label{sec:intro}

Complex (6+ atoms) organic molecules \citep[COMs, ][]{herbst09} are detected in a wide variety of astrophysical environments spanning different stages of star formation: 
from prestellar cores in cold, quiescent dense clouds \citep{marcelino07,bacmann12,cernicharo12,vastel14,izaskun16};  
to luke-warm envelopes \citep{oberg10}, 
hot cores and corinos \citep[see, e.g.,][]{cummins86,blake87,cazaux03,bottinelli04,bianchi19}, 
and shocked outflow regions associated with protostars \citep{arce08,codella17,lefloch18};
and finally in protoplanetary disks around pre-main sequence stars \citep{oberg15,walsh16,jenny18}.  
COMs are also detected in our own Solar System comets \citep[most recently in comet 67P/Churyumov-Gerasimenko,][]{goesmann15,altwegg15}.   
This suggests that planetesimals may have delivered organic precursors of more complex prebiotic species to the early Earth, perhaps contributing to the formation of life. 
%This raises the question on whether they have been inherited from the parental dense cloud, or have been produced \textit{in situ} in the protoplanetary disks. 
%In particular, the detection of COMs in the gas phase of cold environments indicates that cold formation pathways must be relatively efficient.  

%
The two COMs of interest for this paper are NH$_2$CHO (formamide) and chemically similar CH$_3$CHO (acetaldehyde). Both have been detected in comets \citep{bockelee00,crovisier04,biver14,goesmann15}, as well as in many of the above mentioned interstellar and circumstellar environments \citep{rubin71,fourikis74,cazaux03,charnley04,bisschop07,bacmann12,cernicharo12,kahane13,vastel14,lopezsepulcre15,codella15,coutens16,izaskun16,codella17,lee19}.  
NH$_2$CHO, along with its isomer methyl isocyanate (CH$_3$NCO), is the simplest molecule after isocyanic acid (HNCO) that contains the peptide bond ($-$(H$-$)N$-$C($=$O)$-$), the molecular bridge that connects amino acids in proteins. 
Therefore, it has been proposed as a precursor of amino acids and simple proteins \citep{saladino07,saladino12,barone15}.  
%Replacing the amino group (NH$_2-$) for a methyl group (CH$_3-$) leads to acetaldehyde (
CH$_3$CHO %), which is also found in comets \citep{crovisier04,goesmann15},  and 
could also play an important role in astrobiology \citep{hjalmarson01}. 
%In particular, aldehydes are also proposed as precursors to amino acids through the Strecker synthesis \citep[see, e.g.,][]{fresneau15}.  

%cold envelopes, hot corinos and shocks
%So far, though, the presence of NH$_2$CHO in interstellar ices has been only tentatively claimed \citep{raunier04}. 
%As formamide, acetaldehyde has been only tentatively detected in interstellar ices \citep{gibb04}. 
The origin of interstellar NH$_2$CHO and CH$_3$CHO %, either grain surface processes and/or gas-phase reactions, 
is currently unclear \citep[see][for a recent review of the NH$_2$CHO case]{sepulcre19}. 
%The formation of both NH$_2$CHO and CH$_3$CHO i
In the MAGICKAL three-phase chemical model \citep{garrod13} both molecules mainly form through radical-radical reactions between NH$_2^.$ or CH$_3^.$ and HCO$^.$ in icy grain mantles. %NH2+H2CO gas-phase is removed because it has an activation barrier. 
%That reaction is also ruled out by Coutens et al. 2016 because of the different deuterium fraction of NH2CHO and H2CO. They also propose grain-surface chemistry as the most likely origin. However, the different deuterium fractionation are compatible with gas-phase origin according to Skouteris 2017
%According to the observed abundances and rotational temperatures in high-mass star-forming regions, had previously suggested that NH$_2$CHO and likely CH$_3$CHO were mainly produced in the ice mantles, although gas-phase formation of CH$_3$CHO could not be excluded. 
Ice formation pathways was also suggested by observations of abundance patterns in high-mass star forming regions \citep{bisschop07}. 
However, recent observations of low-mass star-forming regions suggest that gas-phase reactions may instead dominate the formation of these species %\citep[at least of NH$_2$CHO,][]
\citep{codella17,skouteris17}. 
%Codella 2017 uses Nahoon, it needs gas-phase NH2+H2CO!!! CH3CHO could be coming from the solid phase
%For example, the observed spatial anticorrelation between NH$_2$CHO and CH$_3$CHO in the L1157-B1 low-mass protostellar shock can only be explained if the formation of NH$_2$CHO is dominated by gas-phase reactions   \citep[while CH$_3$CHO abundance was independent on the formation route,][]{codella17}. 
Taking a middle road, \citet{quenard18} 
%recently stressed the need for a 
proposed that combination of  
gas-phase reactions between NH$_2$ and H$_2$CO\footnote{This gas-phase reaction was not included in \citet{garrod13} because it was thought to have an activation barrier. However, recent works by \citet{barone15} and \citet{skouteris17} support this formation pathway, which is also needed to explain the observed NH$_2$CHO abundances in \citet{codella17}.}, %but removed from MAGICKAL
and solid-state
formation on dust grains (either through hydrogenation of HNCO on the grain surfaces, and/or energetic-processing induced radical-radical %(e.g., NH$_2^.$ and HCO$^.$) 
reactions in the ice mantles) are needed to explain the observed NH$_2$CHO abundances across the different astrophysical environments. %Quenard uses UCLCHEM

One key observation constraining the origin of these species is the 
%relative abundance of NH$_2$CHO with respect to CH$_3$CHO.
NH$_2$CHO/CH$_3$CHO abundance ratio, observed to be $\sim$0.1 in hot corinos around low-mass protostars \citep{lee19} and in the cold envelope of the low-mass protostar IRAS 16293-2422B \citep{jaber14}.
To date, only two studies compare the relative NH$_2$CHO  and CH$_3$CHO formation through ice radical chemistry in comparable ice samples.  
%This has been difficult to reproduce in experiments aimed at simulating  NH$_2$CHO  and CH$_3$CHO formation through ice radical chemistry.
\citet{jones11} and \citet{bennett05} experimentally studied the formation mechanism in electron-irradiated CO:NH$_3$ and CO:CH$_4$ ices, respectively, that could potentially apply to all NH$_2$CHO and CH$_3$CHO formation scenarios involving NH$_3$ and CH$_4$ molecules. 
The proposed initial step is the NH$_3$ N-H or CH$_4$ C-H bond rupture, producing amino or methyl radicals (NH$_2^.$ or CH$_3^.$) and H$^.$ radicals with excess energy. The latter react with CO molecules to produce formyl radicals (HCO$^.$) that recombine with neighboring NH$_2^.$ or CH$_3^.$ to form NH$_2$CHO and CH$_3$CHO.  
They found similar formation rates for both species, so electron or cosmic ray processing of ices cannot easily explain the observed NH$_2$CHO/CH$_3$CHO abundance ratio of $\sim$0.1. %in hot corinos \citep{lee19}\footnote{ A similar ratio is also observed in the cold envelope of IRAS 16293-2422B \citep{jaber14}.}. 
%In addition, while quantum chemical calculations show that NH$_2$CHO could indeed be formed through such route in water-rich amorphous ices \citep[although it competes with reformation of NH$_3$ and CO, the final outcome depending on the relative orientation of the NH$_2^.$ and HCO$^.$ radicals,][]{rimola18},  
%the reformation of CH$_4$ and CO would be the only possible outcome for the reaction of CH$_3^.$ and HCO$^.$ radicals, due to geometrical constraints imposed by the water ice surface \citep{romero16}\footnote{CH$_3$CHO formation is allowed, though, on CO surfaces \citep{lamberts19}.}. 

In space, ice mantles are also exposed to UV photons, especially in %expected to be exposed to UV photons in addition to cosmic rays in different environments: 
the vicinity of massive stars, the surface layers of protoplanetary disks around pre-main sequence stars, the cavities carved out by outflows during the early stages of star formation, and even deep into prestellar cores, where cosmic rays produce secondary UV fields \citep{cecchi92,shen04}. 
%We also note that the main outcome of the UV irradiation of the ice samples is the production of reactive radicals, whose chemistry is then characterized. 
%We expect the same formation mechanisms described in \citet{jones11} and \citet{bennett05} to be involved in our experiments.  
%Some of the results should therefore be generally applicable to all situations where such radicals are generated in astrophysical ices, regardless of the energy source and the composition of the ice. 
%
In this work, we test whether radical ice chemistry in ices exposed to UV radiation can explain the NH$_2$CHO and CH$_3$CHO observations around low-mass protostars. 
We follow the formation of NH$_2$CHO and CH$_3$CHO upon vacuum-ultraviolet (VUV) irradiation of a range of CO:NH$_3$ and CO:CH$_4$ binary ice mixtures to address: 
1) the NH$_2$CHO and CH$_3$CHO formation efficiencies upon VUV photoprocessing of binary ice mixtures, and 
%At the same time, we can compare the percent yields and formation rates upon VUV irradiation of CO:NH$_3$ and CO:CH$_4$ ice samples with those reported for the electron irradiation of the same ices in \citet{jones11} and \citet{bennett05}, since the depostied energies and energy fluxes are comparable. 
2) the kinetics of NH$_2$CHO and CH$_3$CHO formation, including any barriers that could eventually prevent the reaction NH$_2^.$/CH$_3^.$ + HCO$^.$ to take place in interstellar ice mantles.
%
%under different conditions,  simulating the energetic processing of the ice mantles by the secondary UV field also produced in the interior of dense clouds by the interaction of the cosmic rays with the H$_2$ molecules. 
%
The experimental simulations are described in Sect. \ref{sec:exp}. The NH$_2$CHO and CH$_3$CHO conversion yields and formation cross-sections are presented in Sect. \ref{sec:results}, and discussed in Sect. \ref{discussion} in terms of the two questions considered above. Finally, the main conclusions are summarized in Sect. \ref{conclussions}. 

\section{Experimental setup}\label{sec:exp}

\begin{deluxetable*}{cccccccc} 
\tablecaption{Summary of the experiments simulating NH$_2$CHO formation.\label{table_experiments_nh2cho}}
\tablehead{
\colhead{Exp.} & \colhead{Ice comp.} & \colhead{Date} &\colhead{Ratio} & \colhead{Irr. T} & \colhead{Thickness} &  \colhead{N(NH$_2$CHO)$^b$} & \colhead{$\frac{N(NH_2CHO)}{\Delta N(NH_3)}$$^b$}\\
\colhead{} & \colhead{} & \colhead{} & \colhead{} & \colhead{(K)} & \colhead{(ML$^{a}$)} & \colhead{(ML$^{a}$)} & \colhead{(\%)}}% & \colhead{Totally free}} 
%\colnumbers
\startdata
1 & CO:NH$_3$ & 08/2018 & 1:1.2 & - & 202 [20] & - & - \\%& - & - \\%08/17 
\hline
\hline
2 & CO:NH$_3$ & 08/2018 & 2.0:1 & 11 & 176 [18] & 4.5 [1.0] & 19.2 [4.3]\\% & = , 20.5 \\%08/07
3 & CO:NH$_3$ & 02/2019 & 2.1:1 & 10 & 172 [17] & 4.4 [1.0] & 14.7 [3.3]\\% & 3.9, =\\%02/20
\hline
4 & CO:NH$_3$ & 08/2018 & 2.1:1 & 20 & 221 [22] & 6.6 [1.7] & 27.2 [7.1]\\% & = , =\\%08/08
%3$\prime$ & CO:NH$_3$ & 2.24:1 & 18.4 & 176.66 & 10.03 $\times$ 10$^{17}$ & 3.18 [0.34] & 2.60 [0.28] & 10.48 [1.12] \\%02/14
5 & CO:NH$_3$ & 02/2019 & 2.1:1 & 19 & 208 [21] & 5.4 [1.2]  & 18.2 [4.1] \\% & 6.0 , 17.6\\%02/26
%%This experiment has very bad spectra, conversion should be taken with caution since the fit at 6.5 fluence is very different from the fit after 244min of irradiation.
6 & CO:NH$_3$ & 03/2019 & 2.9:1 & 19 & 255 [26] & 6.4 [1.6] & 27.9 [6.8]\\% & = , =\\%03/19
\hline
\hline
%4 & CO:NH$_3$ & 1:1    & 11.4 & 100.87 & 2.51 $\times$ 10$^{17}$ & 1.47 [0.04] & 2.91 [0.09] & 40.38 [1.20] \\%04/26
%4$\prime$ & CO:NH$_3$ & 1:1.21 & 9.5 & 79.69 & 11.13 $\times$ 10$^{17}$ & 1.58 [0.25] & 4.38 [0.69] & 7.40 [1.17] \\%02/15
7 & CO:NH$_3$ & 09/2018 & 1.2:1 & 12 & 231 [23] & 3.7 [0.9] & 17.3 [4.0]\\% & 5.5 , =\\%09/26
8 & CO:NH$_3$ & 02/2019 & 1:1.0 &10 & 179 [18] & 9.9 [2.2] & 21.1 [4.6]\\% & - , =\\%02/25
9 & CO:NH$_3$ & 03/2019 & 1.5:1 & 10 &  247 [25] & 8.1 [1.8] & 20.9 [4.6]\\% & - , =\\%03/11
10 & CO:NH$_3$& 03/2019 & 1.1:1 & 10 & 226 [23] & 8.5 [1.9] & 23.9 [5.2]\\% & 9.2 , =\\%03/13
\hline
%6 & CO:NH$_3$ & 1.19:1 & 11.1 & 411.76 & 6.13 $\times$ 10$^{17}$ & 5.40 [0.24] & 2.41 [0.10] & 17.84 [0.78] \\%10/03
%7 & CO:NH$_3$ & 1.29:1 & 11.6 & 415.08 & 2.41 $\times$ 10$^{17}$ & 4.20 [0.11] & 1.79 [0.04] & 32.51 [0.83] \\%04/16
%8 & CO:NH$_3$ & 1.44:1 & 11.6 & 403.34 & 2.76 $\times$ 10$^{17}$ & 4.01 [0.06] & 1.67 [0.03] & 32.16 [0.51] \\%04/18
%9 & CO:NH$_3$ & 1.36:1 & 20.3 & 419.03& 2.70 $\times$ 10$^{17}$ & 4.28 [0.12] & 1.78 [0.05] & 39.51 [1.11] \\%04/17
%10 & CO$|$NH$_3$ & 1$|$1.06 & 20.3 & 123.53 & 7.24 $\times$ 10$^{17}$ & 0.61 [0.09] & 1.01 [0.15] & 10.05 [1.51] \\%08/22 Flux=5.75e13
11 &CO:NH$_3$ & 03/2019 & 1:1.0 & 19 & 191 [19] & 10.5 [2.5] & 34.9 [8.2]\\% & = , =\\%03/21
\hline
12  & CO:NH$_3$ & 09/2018 & 1.1:1 & 24 & 195 [20] & 4.7 [1.0] &36.8 [7.9] \\% & 6.1 , 34.2 \\%09/28
13  & CO:NH$_3$ & 02/2019 & 1:1.1 & 24 & 159 [16] &6.1 [1.4] & 27.1 [6.3]\\% & = , = \\%02/27
\hline
\hline
14 & CO:NH$_3$ & 08/2018 & 1:2.3 & 12 & 203 [20] & 7.9 [1.7] & 28.4 [6.0] \\% & = , 25.2\\%08/20
15 &CO:NH$_3$ & 03/2019 & 1:3.1 & 10 & 208 [21] & 8.5 [1.9] & 34.2 [7.7] \\%& 10.5 , 30.4\\%03/12 %TOTALLY FREE FIT WITHOUT INITIAL CONDITIONS P0
16 &CO:NH$_3$ &  03/2019 & 1:3.2 & 10 & 222 [22] & 7.9 [1.8] & 25.8 [5.9]\\% & = , 38.5*\\%03/26
\hline
17 & CO:NH$_3$ & 08/2018 & 1:3.2 & 20 & 198 [20] & 7.9 [1.8] & 29.1 [6.7] \\% & = , 24.0\\%08/21
18 & CO:NH$_3$& 03/2019 & 1:2.8 & 19 & 221 [22] & 8.7 [2.0] & 33.9 [8.0] \\% & = , 37.5* \\%03/20 
\hline
\hline
19 & CO:NH$_3$:NH$_2$CHO & 10/2018 & 118:149:1 & - & 281 [28] & - & -\\%10/12 
%  & NH$_3$:NH$_2$CHO &  & - &  & - & - & - & - \\%10/12 JUST IR
%  & NH$_2$CHO & - & - &  & - & - & - & - \\%10/10 11.4K
20 & NH$_2$CHO & 10/2018 & - & - & 35 [4] & - & - \\%& - & - \\%10/10 119K Gaussian fit of the 1328cm-1 band, because the 1388 has a double peak in the pure ice and the standard Gaussian fit for the mixtures does not take that into account. int_tabulated of the 1700cm-1 band leads to N=34.63.  
%  & CO & - & 11.3 &  &  & 0.00 & 0.00 & 0.00 \\%10/17
%  & NH$_3$ & - & 11.5  &  &  & 0.00 & NaN & NaN \\%10/15
%  & NH$_2$CHO & - & 11.2 &  &  &  & NaN & NaN \\%10/16 
\enddata
\tablecomments{
$^{a}$ 1 ML = 10$^{15}$ molecules cm$^{-2}$. Uncertainties include a 20\% error due to the use of band strengths from pure ices (see text).
$^b$ After a fluence of 6.5 $\times$ 10$^{17}$ photons cm$^{-2}$. 
%Values are corrected by the NH2CHO 170to215 ratio of the same day. In any case this is not important since I am using the conversion yield, which is calculated directly from the 170deg spectra. 
%Uncertainties include a 20\% error due to the use of band strengths from pure ices (see text).
%NH3 does not have the 20% error because it is cancelled when substracting two values.
}
\end{deluxetable*}

\begin{deluxetable*}{cccccccc} 
\tablecaption{Summary of the experiments simulating CH$_3$CHO formation.\label{table_experiments_ch3cho}}
\tablehead{
\colhead{Exp.} & \colhead{Ice comp.}& \colhead{Date}  & \colhead{Ratio} & \colhead{Irr. T} & \colhead{Thickness} & \colhead{N(CH$_3$CHO)$^b$} & \colhead{$\frac{N(CH_3CHO)}{\Delta N(CH_4)}$$^b$}\\
\colhead{} & \colhead{} & \colhead{} & \colhead{} & \colhead{(K)} & \colhead{(ML$^{a}$)} & \colhead{(ML$^{a}$)} & \colhead{(\%)} } 
%\colnumbers
\startdata
21 & CO:CH$_4$ & 08/2018  & 1.4:1 & - & 131 [13] & - & - \\%& - & - \\%08/17  
\hline
\hline
22 & CO:CH$_4$ & 05/2018 & 3.4:1 & 12 & 155 [16] & 0.50 [0.11] & 6.1 [1.4] \\%05/31
23 & CO:CH$_4$ & 06/2018 & 2.6:1 & 12 & 200 [20] & 0.57 [0.16] & 4.6 [1.3] \\%06/25
\hline
24 & CO:CH$_4$ & 06/2018 & 4.6:1 & 20 & 167 [17] & 0.16 [0.04] & 2.4 [0.6]\\%06/26
\hline
\hline
25 & CO:CH$_4$ & 05/2018 & 1.5:1 & 12 & 132 [13] & 0.43 [0.10] & 4.3 [1.0]\\%05/17
\hline
26 & CO:CH$_4$ & 05/2018 & 1:1.1 & 21 & 110 [10] & 0.36 [0.10] & 3.2 [0.9]\\%05/30
\hline
\hline
27 & CO:CH$_4$ & 06/2018 & 1:2.1 & 11 & 191 [19] & 0.48 [0.10]$^c$ & 2.3 [0.5]$^c$\\%06/27
\hline
28 & CO:CH$_4$ & 08/2018 & 1:2.4    & 20 & 217 [22] & 0.39 [0.09] & 2.3 [0.5] \\%08/06
\hline
\hline
29 & CO:CH$_4$CH$_3$CHO & 08/2018 & 9.6:6.8:1  & - & 116 [12] &  - & - \\%& - & - \\%08/24 int_tabulated of the 1350 band, because the standard Gaussian fit for the mixtures does not work on the more intense IR bands of the pure ice. 
30 & CH$_3$CHO& 08/2018 & - & - & 47 [5] &  - & - \\%& - & - \\%08/24 int_tabulated of the 1350 band, because the standard Gaussian fit for the mixtures does not work on the more intense IR bands of the pure ice. 
\enddata
\tablecomments{$^{a}$ 1 ML = 10$^{15}$ molecules cm$^{-2}$. Uncertainties include a 20\% error due to the use of band strengths from pure ices (see text). 
$^b$ After a fluence of 6.5 $\times$ 10$^{17}$ photons cm$^{-2}$. %Uncertainties include a 20\% error due to the use of band strengths from pure ices (see text). %NH3 does not have the 20% error because it is cancelled when substracting two values.
$^c$ After a fluence of 5.2 $\times$ 10$^{17}$ photons cm$^{-2}$ due to the lack of good IR spectra for higher fluences.} 
\end{deluxetable*}

The experimental simulations were performed using the SPACE CAT\footnote{Surface Processing Apparatus for Chemical Experimentation to Constrain Astrophysical Theories.} setup located at the Center for Astrophysics. 
This setup consists in an ultra-high vacuum chamber evacuated by a Pfeiffer Turbo HiPace 400 pump
backed by a DUO 10M rotary vane pump to a base pressure of about 5 $\times$ 10$^{-10}$ Torr at room temperature. 
The relevant features for this work are described below, while more information on the setup can be found in \citet{lauck15}. 

\subsection{Ice sample preparation}
The CO:NH$_3$ and CO:CH$_4$ binary ice mixtures were deposited onto a CsI substrate at a temperature of 11 K, achieved by means of a closed-cycle He cryostat. The temperature was monitored by a LakeShore 335 temperature controller with an estimated accuracy of 2 K and a relative uncertainty of 0.1 K. 
CO (gas, 99.95\% purity, Aldrich) and CH$_4$ (gas, 99.9\%, Aldrich) were pre-mixed in a differentially pumped gas line with a base pressure on the order of 10$^{-4}$ Torr and then introduced in the chamber through a 4.8 mm diameter dosing pipe at 17.8 mm from the substrate. 
For the CO:NH$_3$ ice mixtures, CO and NH$_3$ (gas, 99.99\%) were introduced through two different dosing pipes (due to the corrosive nature of NH$_3$) located at 30.5 mm from the substrate. 
In each experiment, the total thickness of the deposited ice samples was $\sim$200 ML (1 ML = 10$^{15}$ molecules cm$^{-2}$). 

We also deposited pure NH$_2$CHO and CH$_3$CHO ices, as well as CO:NH$_3$:NH$_2$CHO and CO:CH$_4$:CH$_3$CHO ternary ice mixtures. 
These ice samples were not photoprocessed, but used to confirm the assignments of the infrared bands and the desorption temperatures of the two main photoproducts. 
To that end, we used NH$_2$CHO (liquid, 99.5\% purity, Sigma) and CH$_3$CHO (liquid, 99.5\% purity, Aldrich) without further purification, after applying two freeze-thaw-pump cycles. 
We note that pure NH$_2$CHO ice was deposited at 119 K, well above the desorption temperatures of N$_2$ and CO$_2$, that appeared as contaminants when depositing NH$_2$CHO at 11 K\footnote{Liquid NH$_2$CHO was warmed up prior to the deposition to increase its vapor pressure in the gas line, leading to the dissociation of some NH$_2$CHO molecules. The dissociation products (including N$_2$ and CO$_2$) were introduced in the chamber along with the intact NH$_2$CHO molecules, and could have become incorporated to the ice samples if the substrate temperature was below their desorption temperature.}. 
%CO2 was introduced as a contaminant during deposition of the ternary mixture, though. I do not remember if liquid NH2CHO was also warmed up prior to this particular deposition. 
In any case, the pure NH$_2$CHO IR spectrum at 119 K shows only subtle differences 
%in the band strengths of some IR bands 
with respect to the one collected at 11 K.
%, and a change in the relative intensities of the double peak structure of the $\sim$1388 cm$^{-1}$ band.  

\begin{figure}
\plotone{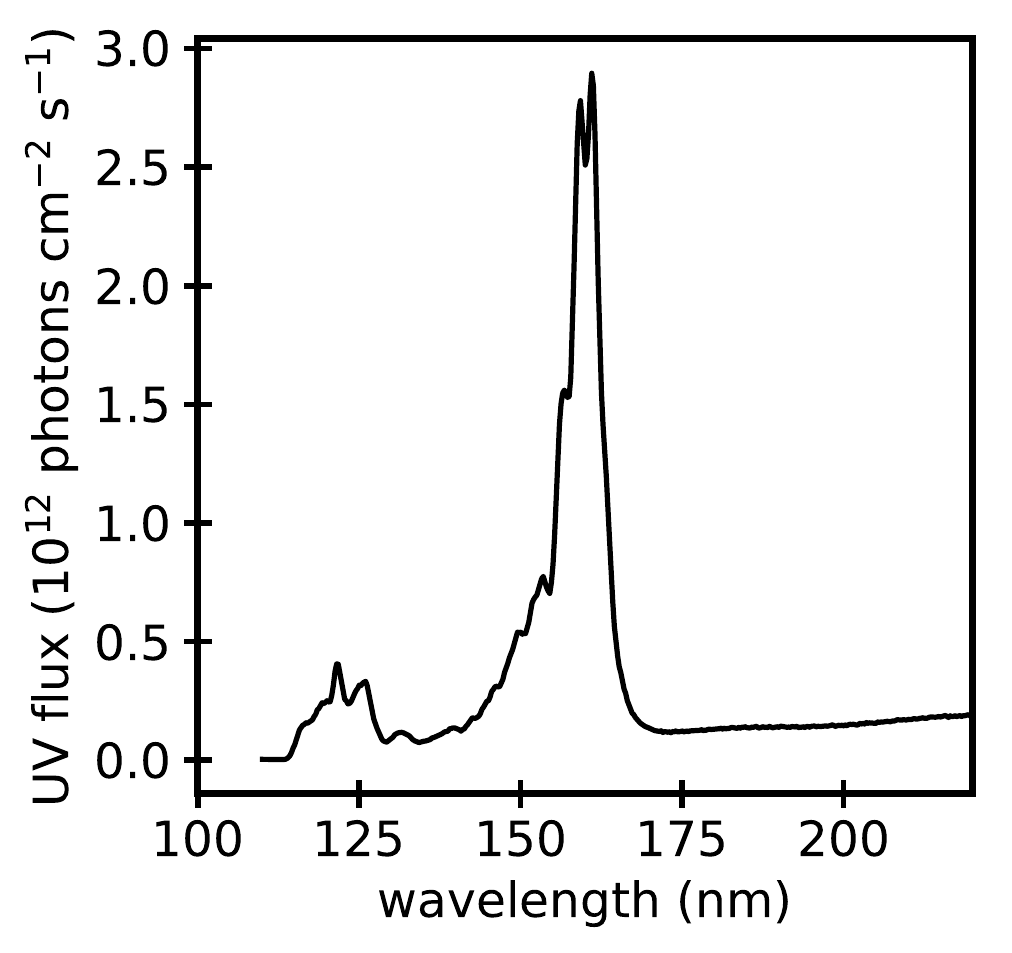}
\caption{Emission spectrum of the Hamamatsu H2D2 L11798  lamp in the VUV range. \label{lampem}}
\end{figure}

\subsection{VUV photoprocessing of the ice samples}
After deposition, the ice samples were VUV irradiated using a deuterium lamp (Hamamatsu H2D2 L11798),  
either at the deposition temperature (11 K) or at 20 K.
The factory lamp emission spectrum in the VUV range (110-220 nm) is presented in Fig. \ref{lampem}, and features a strong emission band at $\sim$160 nm that correspond to the emission of molecular deuterium, and weaker emission ($\sim$30\% of the flux at 160 nm) around Ly-$\alpha$ wavelengths. 
The total VUV flux at the sample position, measured with a NIST calibrated AXUV-100G photo-diode placed in front of the substrate, varied between  
3.7$-$7.4  $\times$ 10$^{13}$ photons cm$^{-2}$ s$^{-1}$. 
Sample irradiation times varied between 210 and 380 min, resulting in total fluences of 6.2$-$12.8 $\times$ 10$^{17}$ photons cm$^{-2}$. 
For comparison, assuming a secondary UV flux of $\sim$10$^4$ photons cm$^{-2}$ s$^{-1}$ in the interior of dense cores \citep{shen04}, interstellar ices are exposed to a fluence of $\sim$3 $\times$ 10$^{17}$ photons cm$^{-2}$ during a cloud lifetime of $\sim$10$^6$ years.

We ran three series of experimental simulations, two for the CO:NH$_3$ (08$-$09/2018 and 02$-$03/2019), and one for the CO:CH$_4$ (05$-$06/2018) ice samples. 
Some differences were found in the NH$_2$CHO formation cross-sections (Sect. \ref{sec:kin}) between experiments of the two series performed under the same conditions, that we think are due to subtle fluctuations in the energy distribution of the VUV photons over time. 
These fluctuations do not affect the COM formation efficiencies in Sect. \ref{sec:conv}, since they are removed when the COMs formation is normalized to the parent molecules dissociation.   
Tables \ref{table_experiments_nh2cho} and \ref{table_experiments_ch3cho} summarize all the experimental simulations presented in this paper, along with the date they were performed. They include the two blank experiments (1 and 21) with no photoprocessing of the binary ice mixtures, used to confirm that NH$_2$CHO and CH$_3$CHO were formed upon VUV irradiation of the ice samples. 

\subsection{IR ice spectroscopy}\label{sec:exp-ir}
During the experimental simulations, the ice samples were monitored with a Fourier transform infrared (FTIR) spectrometer (Bruker Vertex 70v with a liquid-nitrogen-cooled MCT detector) in transmittance. Spectra were collected with a normal IR beam incidence after deposition and after photoprocessing of the binary mixtures, and every 5 minutes during warm-up of the photoprocessed ices (see below); and with a 45$^{\circ}$ incidence every 10 minutes during the photoprocessing of the ice samples; with a resolution of 1 cm$^{-1}$ in the range 4000-400 cm$^{-1}$, and averaged over 128 interferograms. When necessary, the IR spectra were baseline corrected using a \textit{spline} function with the IDL software. The column density of selected species in the ice was calculated from the IR spectra using the equation  

\begin{equation}
N=\frac{1}{A}\int_{band}{\tau_{\nu} \ d\nu},
\label{eqn}
\end{equation}

\noindent where $N$ is the column density in molecules cm$^{-2}$, $\tau_{\nu}$ the optical depth of the absorption band (2.3 times the absorbance), and $A$ the band strength in cm molecule$^{-1}$, and corrected from the 45$^{\circ}$ IR beam incidence, unless otherwise indicated.  
For CO, NH$_3$, and CH$_4$, the integrated optical depth over a particular IR feature was calculated numerically using the function \textit{int\_tabulated} in IDL. % unless otherwise indicated. 
%errors are too small (smaller than the symbol)
%I COULD CALCULATE THE ERROR FROM THE RMS OF THE SPECTRUM AND THE WIDTH OF THE BAND. 
For NH$_2$CHO and CH$_3$CHO, the corresponding IR bands were fitted to a Gaussian using the \textit{curve\_fit} function in Python (see Figures \ref{fit_nh2cho} and \ref{fit_ch3cho}). 
Band strengths are usually measured for pure amorphous ices (Table \ref{strengths}), and the same values are adopted for ice mixtures, which introduces an uncertainty of about 20\% in the absolute column densities \citep{dhendecourt86}\footnote{This uncertainty could be higher in very diluted ice mixtures, not used in this work. For example, a factor of 3 difference is found for the CH$_4$ 1304 cm$^{-1}$ IR band strength in a 20:1 CO:CH$_4$ mixture, compared to the value in a pure amorphous ice \citep{hudgins93}}
%That will be the only source of error for N CO, NH3, CH4 (the error in the Gaussian area for NH3 for example is less than 1%, so it is negligible); while for NH2CHO and CH3CHO it has to be added to the error in the area calculated with the gaussian fit (in the same way that I added a 10% error to the error of the calculated line fluxes with the Gaussian fit too). 
For most of the data analysis we only take into account the statistical errors, however, since the band strength differences between ice experiments of similar composition should be small.
%This 20\% error has not been taken into account when comparing column densities within the same experiment, or between different experiments, since we assume that the same error would apply to all calculated column densities for similar ice mixtures, and only the relative errors were considered in those cases.

\begin{table}[ht!]
\centering
\caption{IR features used to calculate the column density of selected ice components. 
Frequencies and band strengths for pure ices at 10$-$25 K. 
%, except for CH$_3$CHO (theoretical).
\label{strengths}}
\begin{tabular}{ccc}
\hline
\hline
Molecule&Frequency&Band strength\\
&(cm$^{-1}$)&(cm molec$^{-1}$)\\
\hline
CO & 2139 & 1.1$\times 10^{-17}$ $^{a}$\\
NH$_3$ & 1070 & 1.7$\times 10^{-17}$ $^{b}$\\
CH$_{4}$ &1304 & 8.0$\times 10^{-18}$ $^{c}$\\
%NH$_2$CHO & 1328 & 8.5$\times 10^{-18}$ $^{e}$\\ %Only used for pure formamide ice. Not interesting.  
NH$_2$CHO & 1388 & 6.8$\times 10^{-18}$ $^{d}$\\ %It is the best band to follow evolution during TPD
%NH$_2$CHO & 1699 & 6.5$\times 10^{-17}$ $^{d}$\\
%CH$_3$CHO & 1351 & 4.5$\times 10^{-18}$ $^{e}$\\
%CH$_3$CHO & 1426 & 3.6$\times 10^{-18}$ $^{e}$\\
CH$_3$CHO & 1728 & 3.0 $\times 10^{-17}$ $^{e}$\\
\hline
\end{tabular}
\tablecomments{$^{a}$\citet{gerakines95}, $^{b}$\citet{schutte96}, $^{c}$\citet{bouilloud15}, $^{d}$\citet{brucato06}, $^{e}$\citet{hudson20}}.
\end{table}

\subsection{Temperature Programmed Desorption of photoprocessed ice samples}
At the end of the photoprocessing, the ice samples were warmed from the irradiation temperature up to 250 K using a 50 $\omega$ thermofoil heater with a heating rate of 2 K min$^{-1}$, until their complete sublimation was achieved. The IR spectra of the samples were collected every 5 min (10 K), and a quadrupole mass spectrometer (QMS; Pfeiffer QMG 220M1) was used to detect the desorbing molecules in the gas phase. 
The initial ice components, CO, NH$_3$, and CH$_4$, were monitored through the mass fragments $m/z$ = 28 (CO$^{+}$), 17 (NH$_3^+$), and 15 (CH$_3^{+}$, almost the same intensity as the main mass fragment CH$_4^{+}$, shared with atomic oxygen), respectively; and the photoproducts NH$_2$CHO and CH$_3$CHO through the mass fragments $m/z$ = 45 (NH$_2$CHO$^{+}$), and $m/z$ =  29 (HCO$^{+}$, main mass fragment for acetaldehyde), leading to the temperature programmed desorption (TPD) curve for each species.

\section{Results}\label{sec:results}

\subsection{Fiducial CO:NH$_3$ and CO:CH$_4$ VUV irradiation experiments}

Our standard experimental simulations to test the formation of NH$_2$CHO and CH$_3$CHO upon VUV irradiation of CO:NH$_3$ and CO:CH$_4$ ice mixtures, respectively, are 1:1 ice samples irradiated at 10-12 K (Experiments 7-10 for NH$_2$CHO, and Exp. 25 for CH$_3$CHO, see Tables \ref{table_experiments_nh2cho} and \ref{table_experiments_ch3cho}). 

\subsubsection{Formation of NH$_2$CHO}\label{form_nh2cho}

\begin{figure*}
\centering
\plotone{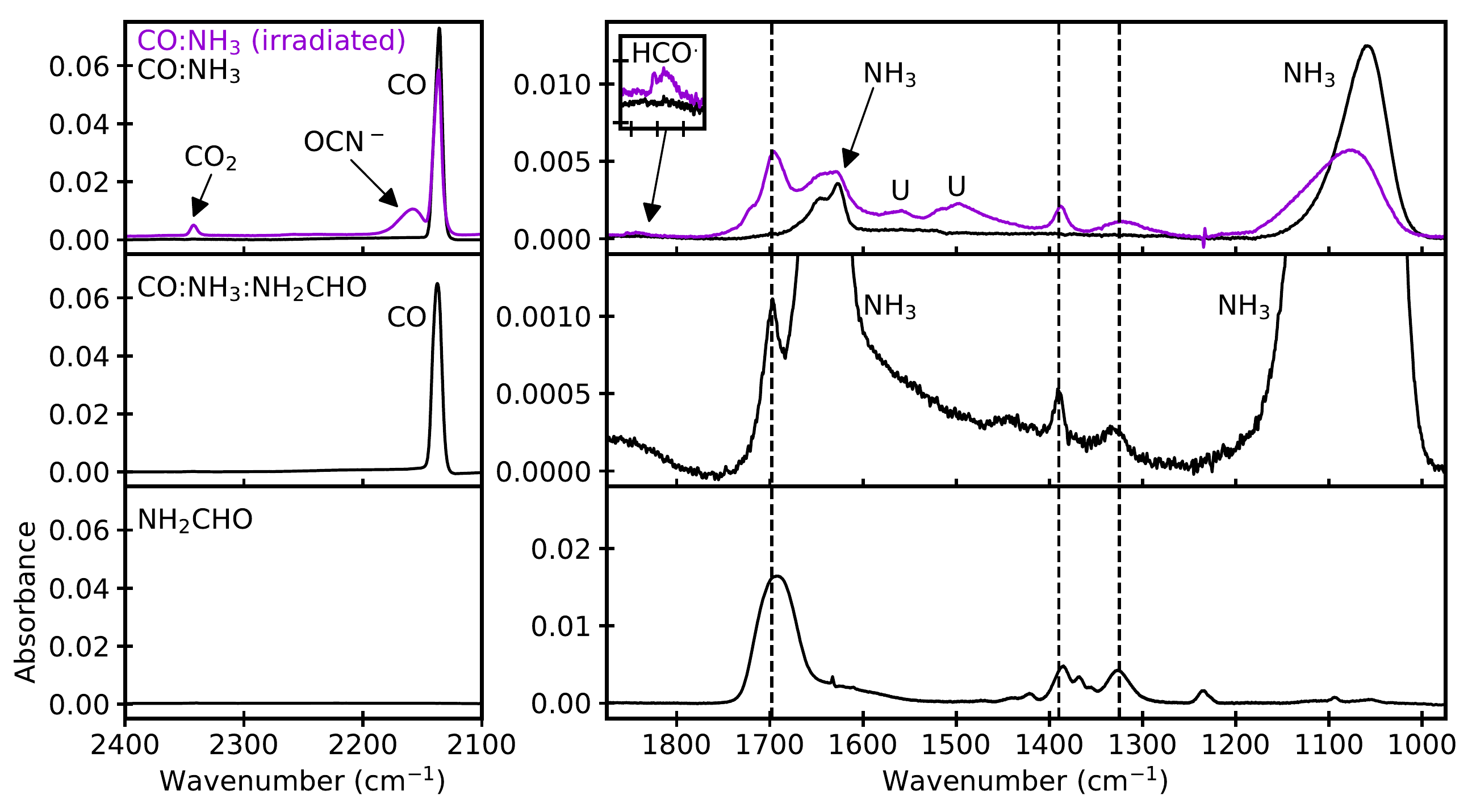}
\caption{ 
\textit{Top panels}: IR spectrum of a CO:NH$_3$ 1:1 ice sample before (solid black) and after (solid violet) VUV photoprocessing at 10 K (Exp. 10). 
\textit{Middle panels}: IR spectrum of a CO:NH$_3$:NH$_2$CHO 118:149:1 ice sample at 12 K (Exp. 19). 
\textit{Bottom panels}: IR spectrum of a pure NH$_2$CHO ice deposited at 119 K (Exp. 20). 
IR band assignments are indicated in the panels (U corresponds to unassigned features). 
Vertical dashed lines indicate the positions of the three most intense NH$_2$CHO IR bands. 
%The intensity of the C=O stretching band (2135 cm$^{-1}$) and the NH$_3$ umbrella band (1060 cm$^{-1}$) decrease as the result of photodestruction, while new IR features are observed due to the formation of photoproducts. 
%In particular, dashed lines indicate the position of the formamide ice IR features close to the NH$_3$ deformation band according to \citet{brucato06} (see Fig. \ref{irspecnh2cho}). %at 1708 cm$^{-1}$, 1387 cm$^{-1}$, and 1328 cm$^{-1}$ . 
\label{irnh2cho}}
\end{figure*}

\begin{figure}
    \centering
    \includegraphics[width=6.9cm]{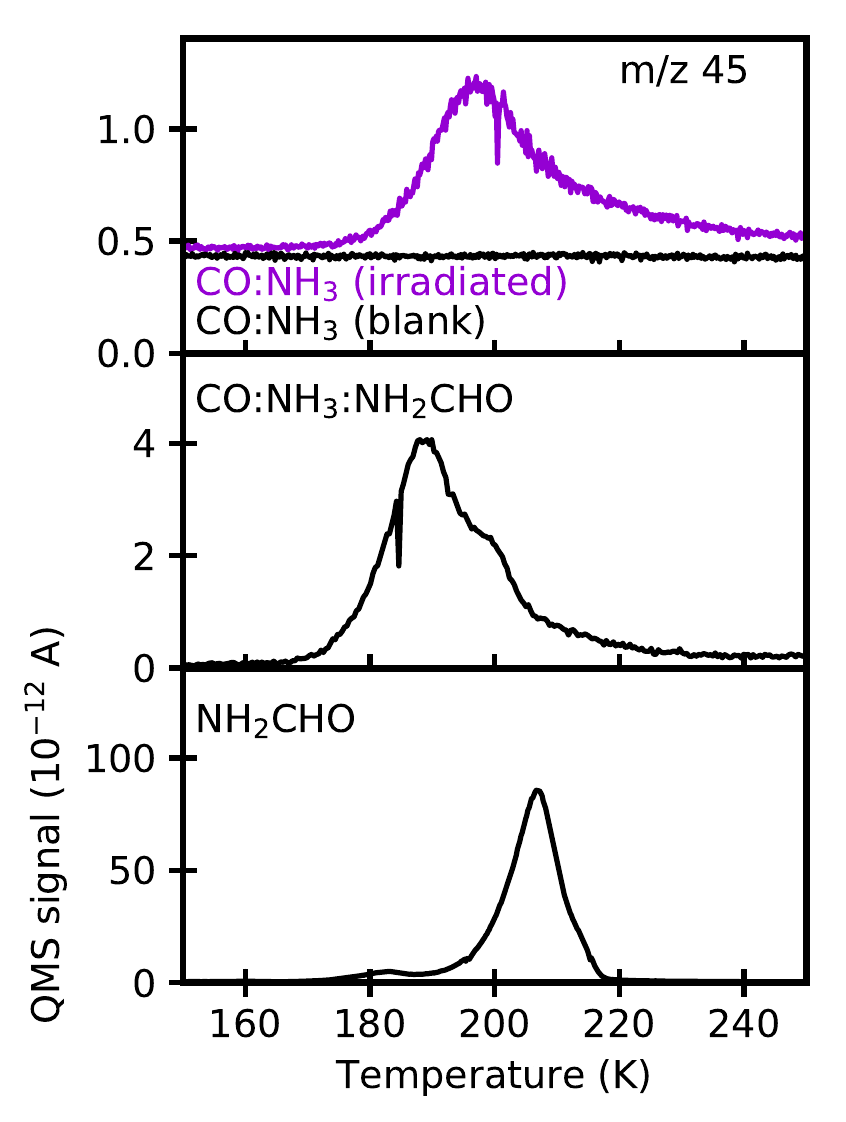}
    \caption{The TPD curve of the m/z = 45 fragment corresponding to the produced NH$_2$CHO after photoprocessing of a CO:NH$_3$ ice sample (\textit{top panel}, solid violet) is compared to the thermal desorption of NH$_2$CHO in a CO:NH$_3$:NH$_2$CHO ice mixture (\textit{middle panel}), and a pure NH$_2$CHO ice (\textit{bottom panel}). No NH$_2$CHO desorption was detected for the blank experiment with no photoprocessing of the CO:NH$_3$ ice sample (\textit{top panel}, solid black). \label{tpdnh2cho}}
   
\end{figure}

The top panels of Fig. \ref{irnh2cho} show the IR spectrum in the 2400$-$975 cm$^{-1}$ range of the fiducial CO:NH$_3$ 1:1 ice mixture (Exp. 10) before irradiation (black), and after a fluence of 6.5 $\times$ 10$^{17}$ photons cm$^{-2}$ (violet).  
Photodestruction of CO and NH$_3$ is observed by the decrease of the C=O stretching band peaking at 2135 cm$^{-1}$ (left panel), and the NH$_3$ umbrella band at 1060 cm$^{-1}$ (right panel). 
At the same time, formation of NH$_2$CHO is evidenced by the growth of IR bands at 1698 cm$^{-1}$, 1390 cm$^{-1}$, and 1325 cm$^{-1}$, which correspond to the position of NH$_2$CHO IR bands in a CO:NH$_3$:NH$_2$CHO ice mixture (middle panels), and in a pure NH$_2$CHO ice (bottom panels). 
%The weak shoulder in the formamide C=O stretching band (1698 cm$^{-1}$) peaking at 1722 cm$^{-1}$ probably corresponds to H$_2$CO. 
Small features due to the formation of CO$_2$ (2342 cm$^{-1}$), OCN$^{-}$ (2160 cm$^{-1}$), and unreacted HCO$^{.}$ radicals (1850 cm$^{-1}$) are also observed.  
Two IR features peaking at 1561 cm$^{-1}$  %glutamic acid has an IR band in this range. However, I think it is too complex to be produced and lead to such an intense IR band (it was not monitored with the QMS)
%$\sim$1524 cm$^{-1}$, 
and 1496 cm$^{-1}$ remained unassigned. %There could be a contribution of the H2CO band at 1494 cm-1, but it is not the main carrier of this band, since it is more intense than the 1720cm-1 shoulder. 

The top panel of Fig. \ref{tpdnh2cho} shows the TPD curve of NH$_2$CHO ($m/z$ = 45) after photoprocessing of the CO:NH$_3$ ice sample (violet). Thermal desorption of the photoproduced NH$_2$CHO peaks at a temperature of 197 K, in between the desorption temperature of a pure NH$_2$CHO ice (207 K, bottom panel) and NH$_2$CHO in a CO:NH$_3$:NH$_2$CHO ice mixture (189 K, middle panel). 
Desorption was also observed at the same temperature for the mass fragments $m/z$ = 44 (NHCHO$^+$) and $m/z$ = 29 (HCO$^+$), shown in the Appendix. 
Since the intensity of the $m/z$ = 44 desorption peak is lower than that of the $m/z$ = 45 peak, even if some trapped CO$_2$ molecules are contributing to the former, the potential contribution of $^{13}$CO$_2$ molecules to the latter would be negligible.
No $m/z$ = 45 signal was detected in the blank experiment with no irradiation of the CO:NH$_3$ ice sample. 
Similar results were found for the rest of the experiments in Table \ref{table_experiments_nh2cho}, i.e., NH$_2$CHO is formed in all irradiated CO:NH$_3$ ice experiments demonstrating that NH$_2$CHO formation in cold CO:NH$_3$ ices is robust. 

\subsubsection{Formation of CH$_3$CHO}\label{form_ch3cho}

\begin{figure*}
\centering
\plotone{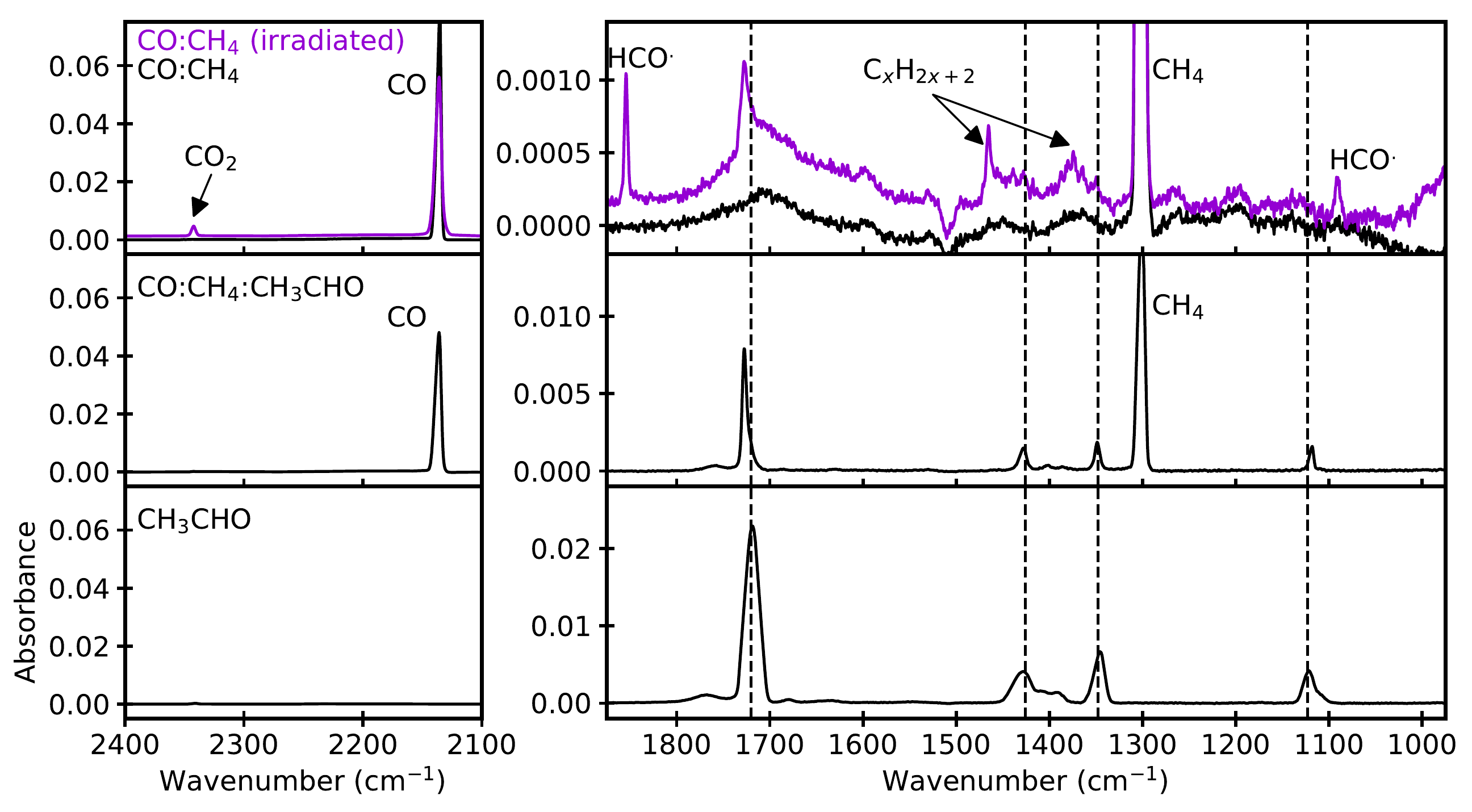}
\caption{ 
\textit{Top panels}: IR spectrum of a CO:CH$_4$ ice sample before (solid black) and after (solid violet) VUV photoprocessing (Exp. 25). 
%The intensity of the C=O stretching band (2135 cm$^{-1}$) and the CH$_4$ deformation band (1304 cm$^{-1}$) decrease as the result of photodestruction, while new IR features are observed due to the formation of photoproducts. In particular, 
\textit{Middle panels}: IR spectrum of a CO:CH$_4$:CH$_3$CHO 9.6:8.4:1 ice sample at 11 K (Exp. 29). 
\textit{Bottom panels}: IR spectrum of a pure CH$_3$CHO ice at 11 K (Exp. 30). 
IR band assignments are indicated in the panels. 
Vertical dashed lines indicate the position of the IR features of pure acetaldehyde at 1720 cm$^{-1}$, 1426 cm$^{-1}$, and 1348 cm$^{-1}$.  
%the bands at $\sim$1426 cm$^{-1}$ and $\sim$1351 cm$^{-1}$ correspond to the $\nu_{12}$ and $\nu_{7}$ CH$_3$- deformation vibrational modes in acetaldehyde \citep{bennett05}. 
\label{irch3cho}}
\end{figure*}

\begin{figure}
    \centering
    \includegraphics[width=6.9cm]{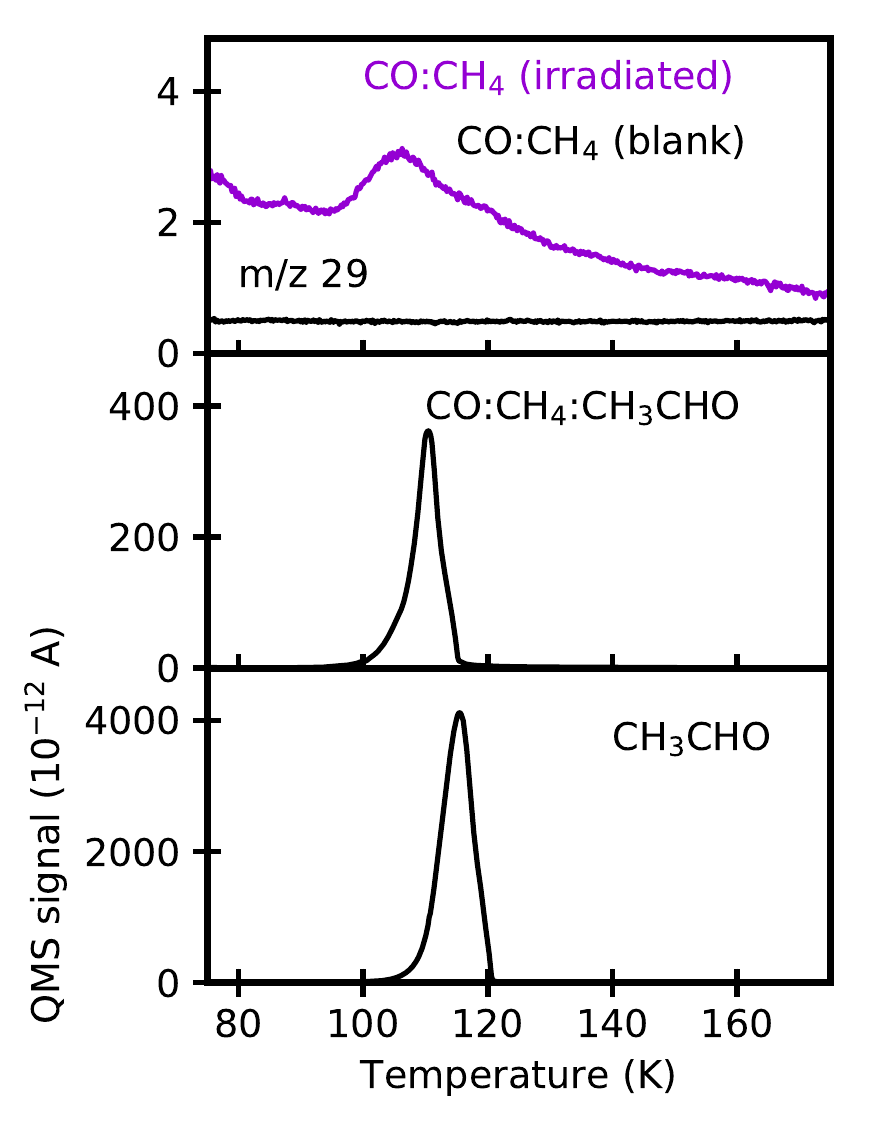}
    \caption{The TPD curve of the m/z = 29 (HCO$^+$) fragment used to verify the CH$_3$CHO production after photoprocessing of a CO:CH$_4$ ice sample (\textit{top panel}, solid violet) is compared to the thermal desorption of CH$_3$CHO in a CO:CH$_4$:CH$_3$CHO ice mixture (\textit{middle panel}), and a pure CH$_3$CHO ice (\textit{bottom panel}). No CH$_3$CHO desorption was detected for the blank experiment (\textit{top panel}, solid black)    \label{tpdch3cho}}
\end{figure}

The top panels of Fig. \ref{irch3cho} show the IR spectrum in the 2400$-$975 cm$^{-1}$ range of the CO:CH$_4$ ice mixture in Exp. 25 before irradiation (black) and after a total fluence of 8.8 $\times$ 10$^{17}$ photons cm$^{-2}$ (violet). 
A new IR feature due to the formation of CH$_3$CHO appears at 1728 cm$^{-1}$. This feature is slightly blueshifted in mixtures with CO and CH$_4$ (middle panels) with respect to its position in a pure CH$_3$CHO ice (bottom panels). The other CH$_3$CHO IR bands at 1426 cm$^{-1}$, 1348 cm$^{-1}$, and 1123 cm$^{-1}$ could not be unambiguously detected in the photoprocessed ice. 
IR bands corresponding to the intermediate photoproduct HCO$^{.}$ are observed at 1855 cm$^{-1}$ and 1092 cm$^{-1}$, and also the CO$_2$ feature at 2342 cm$^{-1}$. 
The band peaking at 1465 cm$^{-1}$ corresponds to the -CH$_2$- scissors and/or the CH$_3$- antisymmetric deformation vibrational modes in saturated hydrocarbons (C$_x$H$_{2x+2}$), that are also produced during irradiation of CO:CH$_4$ ice mixtures, while the CH$_3$- symmetric deformation vibrational mode could contribute to the band at 1377 cm$^{-1}$. %According to Mahesh, it could also be C2H4O (m/z=44), but anyways hydrocarbons have been detected with the QMS too
For example, additional C$_2$H$_6$ IR features are detected at 2972 cm$^{-1}$ and 2980 cm$^{-1}$ (not shown in Fig. \ref{irch3cho}). 

The TPD curve of HCO$^{+}$ (m/z = 29) corresponding to CH$_3$CHO molecules after photoprocessing of the ice sample in Exp. 24 is presented in Fig. \ref{tpdch3cho}. The photoproduced acetaldehyde thermal desorption peaks at 106 K, similar to a pure acetaldehyde ice (115 K, bottom panel), and acetaldehyde in a CO:CH$_4$ matrix (110 K, middle panel). 
Desorption was also observed at the same temperature for the mass fragment $m/z$ = 15 (CH$_3^+$, shown in the Appendix), and tentatively detected for the mass fragments $m/z$ = 44 (CH$_3$CHO$^+$) and $m/z$ = 43 (CH$_2$CHO$^+$)  (not shown). 
CO desorption was not detected in this temperature range (see Appendix), ruling out any $^{13}$CO contribution to the $m/z$ = 29 desorption peak.
No $m/z$ = 29 signal was detected for the blank experiment with no irradiation of the ice mixture. 
Results were similar for the rest of experiments in Table \ref{table_experiments_ch3cho}.

\subsection{NH$_2$CHO and CH$_3$CHO formation efficiencies}\label{sec:conv}

\begin{figure}
\centering
\includegraphics[width=6.5cm]{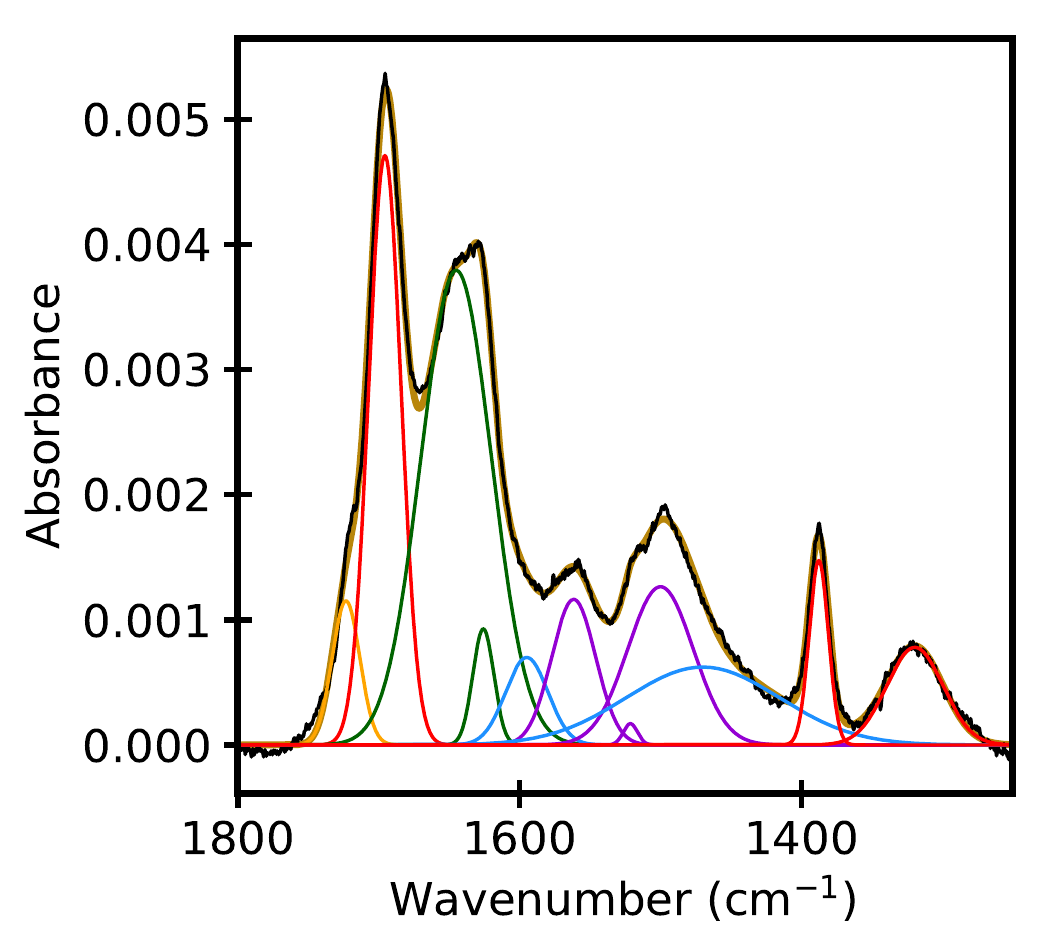}
\caption{Example fit to assigned and unassigned features in the IR spectrum of a VUV photoprocessed CO:NH$_3$ ice sample (Exp. 9). The measured IR spectrum is shown in black with the total fit overimposed in gold. Eleven Gaussians were used for the fit (see text), including three corresponding to the NH$_2$CHO IR features at 1698 cm$^{-1}$, 1390 cm$^{-1}$, and 1325 cm$^{-1}$ (red), one probably corresponding to H$_2$CO (orange), two corresponding to the NH$_3$ deformation mode (green), three for the unassigned features (purple), and two extra Gaussians to better reproduce the shape of the spectrum (blue). 
\label{fit_nh2cho}}
\end{figure}

\begin{figure}
\plotone{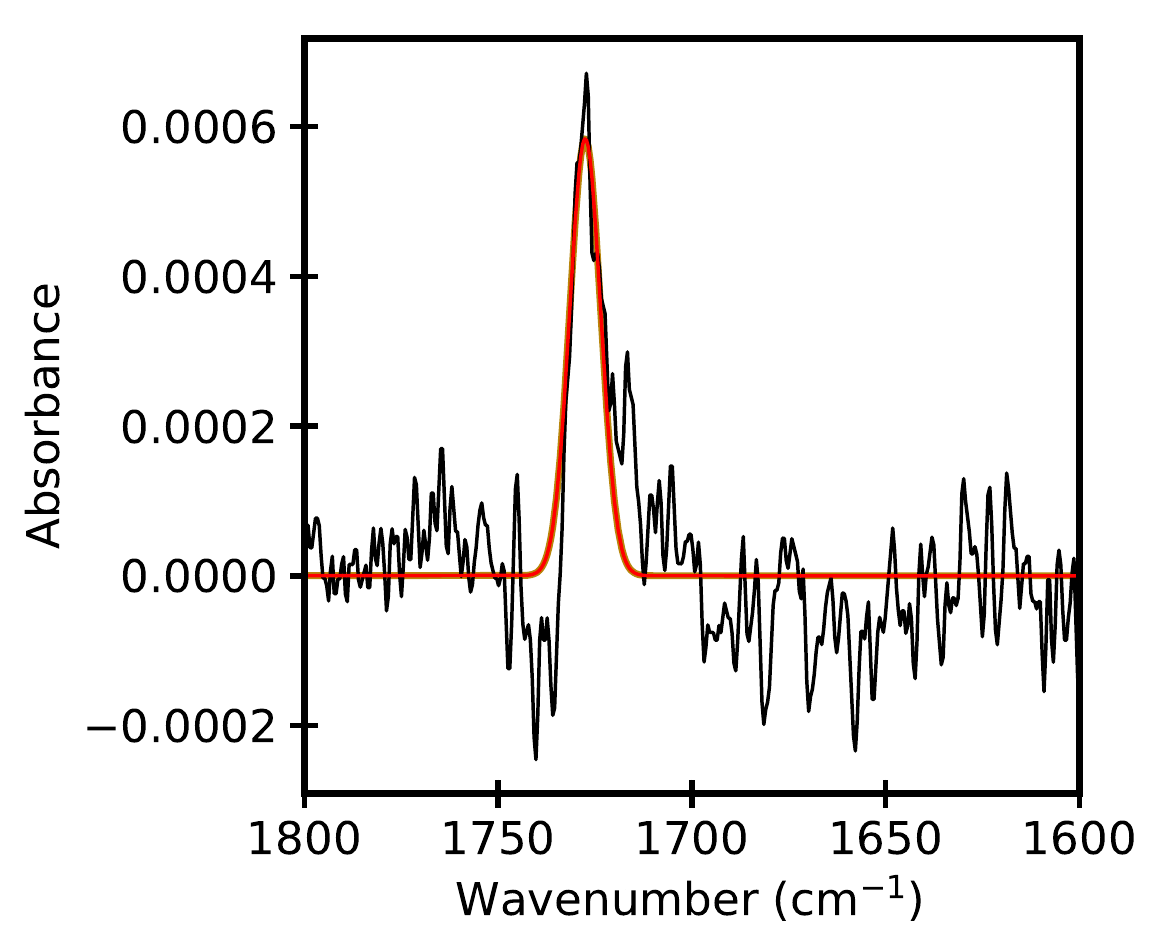}
\caption{Example fit to the 1728 cm$^{-1}$ IR feature appeared in the IR spectrum of a CO:CH$_4$ ice mixture at the end of the irradiation in Exp. 24. The measured IR spectrum is shown in black with the fit in red. 
%Nine Gaussians were used for the fit (blue), including two corresponding to the IR features of CH$_3$CHO at 1427 cm$^{-1}$ and 1350 cm$^{-1}$ (red). The latter was used to extract the acetaldehyde column densities, since it appears in a cleaner region of the spectrum, and has a stronger intrinsic intensity. 
\label{fit_ch3cho}}
\end{figure}

To quantify the formation efficiencies of NH$_2$CHO and CH$_3$CHO in photoprocessed CO:NH$_3$ and CO:CH$_4$ ices, we define the conversion yield as the amount of photodissociated NH$_3$ or CH$_4$ that is converted into  NH$_2$CHO and CH$_3$CHO after a given fluence (we have used 6.5 $\times$ 10$^{17}$ photons cm$^{-2}$ unless otherwise indicated). 
This enables us to remove the effects of VUV lamp energy distribution fluctuations between experiments, %flux does not affect since we are calculating the conversion yield at a given fluence
temperature-dependent effective photodissociation rates, 
as well as the different NH$_3$ and CH$_4$ photodissociation cross-sections \citep{gustavo14a,gustavo14b},
when comparing COM formation in experiments with different ice compositions and irradiation temperatures (see Tables \ref{table_experiments_nh2cho} and \ref{table_experiments_ch3cho}). 
Assuming that photodissociation (photodesorption can be considered negligible compared to photodissociation in thick ices) of NH$_3$ and CH$_4$ mainly produces NH$_2^.$ \citep[as a first approximation, see, e.g.,][and references therein]{md17} and CH$_3^.$ \citep{oberg10b}\footnote{Since some photodoissociation events could lead to the production of  CH$_2^.$, the conversions presented here should be treated as upper limits. 
\citet{oberg10b} estimated a CH$_3^.$:CH$_2^.$ branching ratio $\ge$3:1 for the photoprocessing of a pure CH$_4$ ice with a microwave-discharged hydrogen-flow lamp peaking at Ly-$\alpha$, but found that this ratio was even higher in binary mixtures with H$_2$O.}, this conversion yield describes the fraction of photoproduced radicals that react to form NH$_2$CHO and CH$_3$CHO compared to other outcomes, i.e. it is effectively a product branching ratio measure.

The NH$_3$ and CH$_4$ column densities were estimated from the umbrella IR band at 1060 cm$^{-1}$, and the CH$_4$ deformation band at 1304 cm$^{-1}$, respectively, using Eq. \ref{eqn}. The corresponding band strengths are listed in Table \ref{strengths}, and the integrated absorbances were calculated using the function \textit{int\_tabulated} in IDL (Sect. \ref{sec:exp-ir}). 

The production of NH$_2$CHO was estimated from the 1390 cm$^{-1}$ IR band, since it is the least blended band in our experiments. 
%presented the smaller fitting errors among the two unblended bands at 1390 cm$^{-1}$ and 1325 cm$^{-1}$.
To isolate the 1390 cm$^{-1}$ IR band, the IR spectra collected during photoprocessing of the ice samples were fitted in the 1800$-$1250 cm$^{-1}$ range with eleven different Gaussians (Fig. \ref{fit_nh2cho}), including 
three corresponding to the three NH$_2$CHO IR features at 1698 cm$^{-1}$, 1390 cm$^{-1}$, and 1325 cm$^{-1}$ (red); 
two for the doubled-peak NH$_3$ deformation band at $\sim$1642 cm$^{-1}$ and $\sim$1626 cm$^{-1}$ (green); 
and three Gaussians for the two unassigned bands peaking at $\sim$1561 cm$^{-1}$ and $\sim$1496 cm$^{-1}$ (purple). 
In addition, we used one additional Gaussian for the weak shoulder peaking at $\sim$1722 cm$^{-1}$ in the NH$_2$CHO C=O stretching band (1698 cm$^{-1}$), probably corresponding to H$_2$CO (orange).  
%The IR bands at 1625 cm$^{-1}$ and 1593 cm$^{-1}$ could actually correspond to urea \citep{ligterink18}
One Gaussian centered at $\sim$1593 cm$^{-1}$, and one broader Gaussian centered at $\sim$1435 cm$^{-1}$ were also included in the fit to better reproduce the shape of the spectrum (blue). 
The central position of the Gaussians was one of the free parameters used during the fits, and typically varied within 3 cm$^{-1}$ with respect to the above positions from one spectrum to another.
%The 1700cm-1 and the 1388cm-1 bands behave the same in the 2:1 mixture. For the 1:1 and 1:3, it seems that the 1700cm-1 band could be affected by NH3 band, leading to an underestimation of the column density (show one Gaussian fit for each composition). 

%The uncertainties in the calculated column densities include a 20\% error in addition to the fit errors, due to the use in ice mixtures of pure ice band strengths (see Sect. \ref{sec:exp-ir}). 
The production of CH$_3$CHO was estimated from the C=O stretching IR band at 1728 cm$^{-1}$. 
This band could in theory include a contribution from H$_2$CO, but since no H$_2$CO signal was observed during the TPD of the photoprocessed CO:CH$_4$ ices, we assume that CH$_3$CHO is the only contributor to the 1728 cm$^{-1}$ band. 
%Even though this feature is intrinsically stronger than the one used to estimate formamide column density (see Table \ref{strengths}), a lower acetaldehyde column density formed in these experiments led to a lower signal-to-noise ratio in the collected spectra (Fig. \ref{fit_ch3cho}).  
This feature was typically weak in our experiments, and the line intensity was extracted using a single Gaussian fit (Fig. \ref{fit_ch3cho}).

\begin{figure}
\centering
\includegraphics[width=7.75cm]{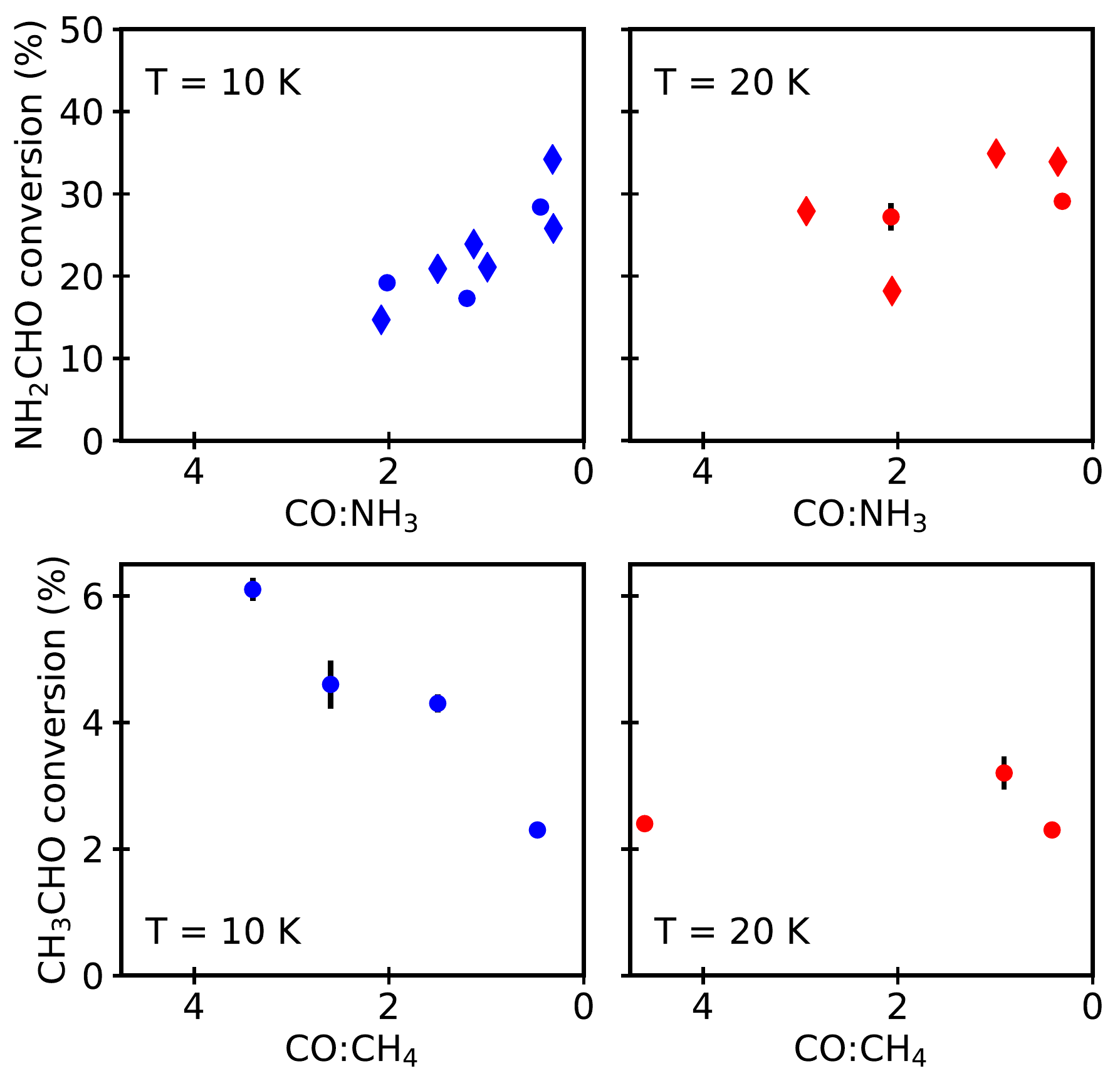}
\caption{Conversion from NH$_2^.$ or CH$_3^.$ radicals into NH$_2$CHO or CH$_3$CHO after irradiation of CO:NH$_3$ or CO:CH$_4$ ice mixtures of different compositions (\textit{top} and \textit{bottom} panels, respectively) with 6.5 $\times$ 10$^{17}$ photons cm$^{-2}$, performed at T = 10 K (\textit{left panels}) and T = 20 K (\textit{right) panels}.
The conversion yields are listed in the last column of Tables \ref{table_experiments_nh2cho} and \ref{table_experiments_ch3cho}, and represented by blue and red markers for the experimental simulations performed at T $\sim$ 10 K (\textit{left} panels) and T = 20$-$24 K (\textit{right} panels), respectively; with error bars in black (without taking into account the 20\% uncertainty due to the use of pure ice band strengths in ice mixtures, see Sect. \ref{sec:exp-ir}). 
We have used dots and diamonds for the first and the second series of CO:NH$_3$ experiments, respectively.
\label{fig_conversion}}

\end{figure}

\begin{figure}
\centering
\includegraphics[width=8cm]{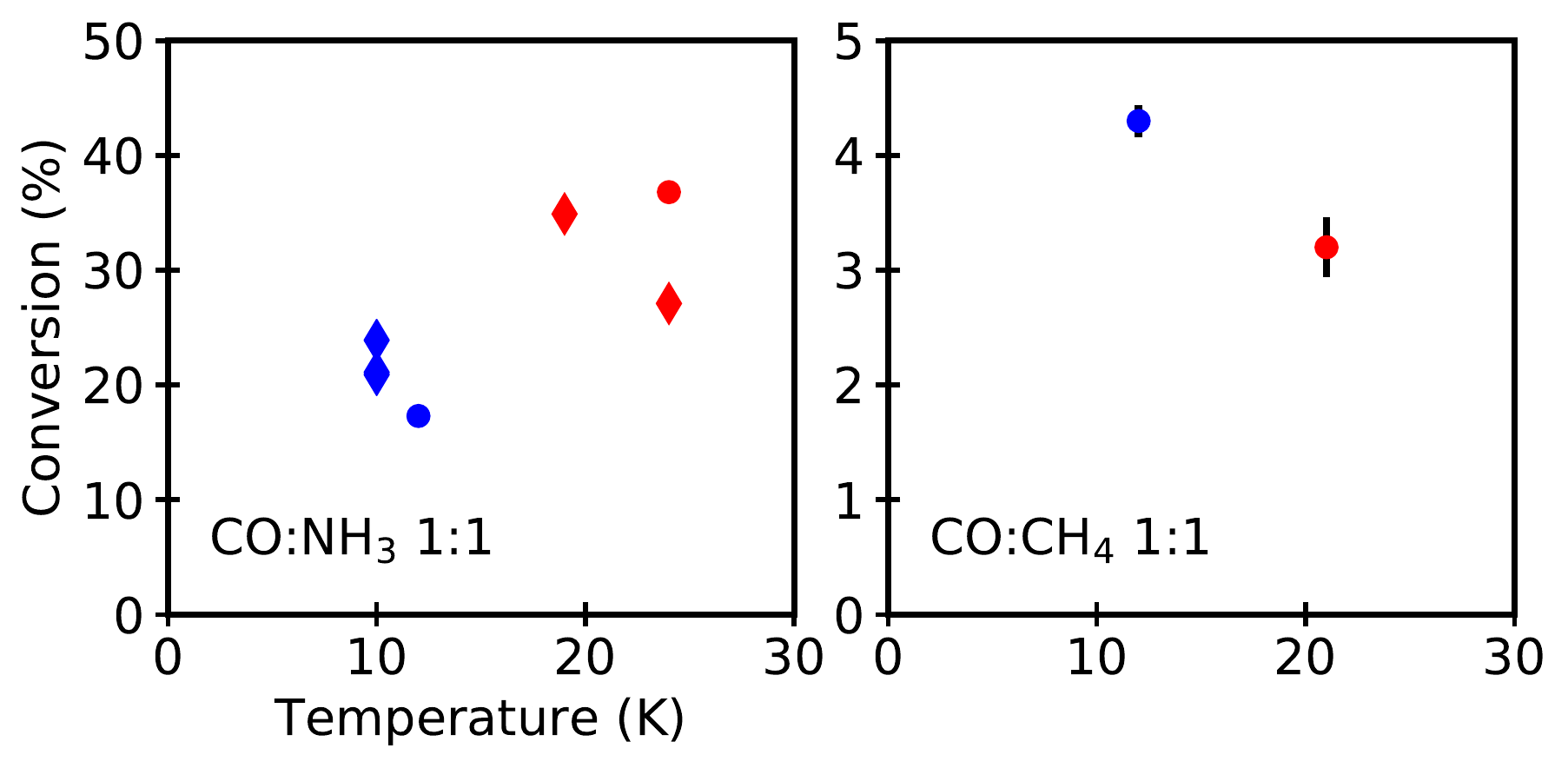}
\caption{Conversion from NH$_2^.$ or CH$_3^.$ radicals into NH$_2$CHO or CH$_3$CHO after irradiation of 1:1 CO:NH$_3$ or CO:CH$_4$ ice samples (\textit{left} and \textit{right} panel, respectively) at T $\sim$ 10 K (blue markers) and T = 20$-$24 K (red markers) with 6.5 $\times$ 10$^{17}$ photons cm$^{-2}$.  Error bars in black do not take into account the 20\% uncertainty due to the use of pure ice band strengths in ice mixtures (see Sect. \ref{sec:exp-ir}). 
We have used dots and diamonds for the first and the second series of CO:NH$_3$ experiments, respectively. 
\label{fig_conversion_1_1}} 
\end{figure}

The conversion yield after a fluence of 6.5 $\times$ 10$^{17}$ photons cm$^{-2}$ for the fiducial CO:NH$_3$ 1:1 ice mixtures photoprocessed at 10 K (Exp. 7-10) is 17-24\% (see last column of Table \ref{table_experiments_nh2cho}), %75% of OCN-, radicals that react at higher temperatures, and maybe unassigned features. Production of N2, N2H2 and N2H4 cannot be confirmed with the TPD. 
while the conversion yield for the fiducial CO:CH$_4$ 1:1 ice mixture VUV irradiated at 12 K (Exp. 25) is $\sim$4\%  (see last column of Table \ref{table_experiments_ch3cho}). 
The conversion from NH$_2^.$ to NH$_2$CHO is, therefore, $\sim$6 times higher than the conversion from CH$_3^.$ to CH$_3$CHO under similar experimental conditions.

In addition to our standard experimental simulations, we carried out experiments with different ice compositions and irradiation temperatures to study their influence in the COM formation efficiencies and kinetics.
The conversion yields after a fluence of 6.5 $\times$ 10$^{17}$ photons cm$^{-2}$ are presented in the last columns of Tables \ref{table_experiments_nh2cho} and \ref{table_experiments_ch3cho}, and shown in Fig. \ref{fig_conversion}. They range from 15\% to 37\%, approximately, for the case of NH$_2$CHO,  and 2$-$6\% for CH$_3$CHO, depending on the ice composition and the irradiation temperature. 
The conversion from NH$_2^.$ to NH$_2$CHO is up to one order of magnitude higher than the conversion from CH$_3^.$ to CH$_3$CHO under the different experimental conditions explored in this work. 

%\subsubsection{Composition and temperature dependence}

%When comparing the formation of NH$_2$CHO or CH$_3$CHO in ices with different compositions, 
%we observe 
The left top panel in Fig. \ref{fig_conversion} show that the conversion from  NH$_2^.$ radicals into NH$_2$CHO is slightly dependent on the ice composition at low temperatures (T $\sim$ 10 K), with higher conversions for the NH$_3$-rich ices: an increase in NH$_3$ concentration by a factor of 6 produces an increase in conversion yield of a factor of 2; 
whereas 
%the dependence on composition is negligible 
no clear trend with the composition is observed 
when the ice is irradiated at T $\sim$ 20 K (upper right panel in Fig. \ref{fig_conversion}). %%Say that the dependence cannot be confirmed instead of negligible
On the other hand, the conversion yield from CH$_3^.$ radicals into CH$_3$CHO \textit{decreases} with CH$_4$ concentration at T $\sim$ 10 K 
(a factor of $\sim$3 variation for a factor of $\sim$7 change in the CO:CH$_4$ ratio, bottom left panel of Fig. \ref{fig_conversion}),  
while the CH$_3$CHO conversion yield seems to be independent of the composition at T = 20 K%, similarly to the NH$_2$CHO production 
(bottom rigth panel of Fig. \ref{fig_conversion}). %20K is the onset of the CO desorption (see fig in 20180807) but at 24K the QMS signal is still 1000 times lower than at the desorption peak. 

Regarding the COMs formation at different temperatures, 
the NH$_2$CHO conversion yield is observed to increase with temperature, but only in ices with CO:NH$_3$ ratio $\ge$1 (top panels in Fig. \ref{fig_conversion}).  %and left panel of Fig. \ref{fig_conversion_1_1}),  
Conversely, 
%the average CH$_3$CHO conversion yield at T = 20 K ($\sim$2\%) is similar to the lowest yield measured at T $\sim$ 10 K in the most concentrated CH$_4$ ices (bottom panels in Fig. \ref{fig_conversion}), i.e., 
the CH$_3$CHO formation efficiency decreases with temperature in ices with CO:CH$_4$ ratio $\ge$1. 
Figure \ref{fig_conversion_1_1} highlights the opposing temperature trends observed for NH$_2$CHO and CH$_3$CHO conversion yields.  

\subsection{NH$_2$CHO and CH$_3$CHO formation kinetics}\label{sec:kin}

\begin{deluxetable}{cccc} 
\tablecaption{Summary of the best-fit NH$_2$CHO formation cross-sections\label{table_kineticsnh2cho}}
\tablehead{
\colhead{Exp.} & \colhead{Ratio} & \colhead{Irr. T} & \colhead{\textit{$\sigma_{form}$}(NH$_2$CHO)}\\% & \colhead{$\sigma_{dest}$(NH$_3$) } & $\sigma_{form}/\sigma_{dest}$\\
\colhead{} & \colhead{} & \colhead{(K)} & \multicolumn{1}{c}{($\times$ 10$^{-18}$ cm$^{2}$)}} 
%\colnumbers
\startdata
2 & 2.0:1 & 11 & 3.0 [0.2]\\% & 3.07 [0.57] & 0.96 [0.19] \\%08/07
3 & 2.1:1 & 10 & 6.3 [0.2]\\%& 4.10 [0.46] & 1.56 [0.19] \\%02/20
\hline
4& 2.1:1 & 20 & 3.5 [0.4]\\% &2.00 [0.41] & 1.76 [0.41] \\%08/08 
%3$\prime$ & 2.24:1 & 18.4 & 176.66 & 5.57 $\times$ 10$^{13}$ & 10.03 $\times$ 10$^{17}$ & 3.18 [0.34] & 3.32 [0.05] & 9.66 [0.64] & 5.38 [0.36] \\%02/14
5 & 2.1:1 & 19 &4.1 [0.2]\\% &7.17 [1.28] & 0.58 [0.12]\\%02/26 a F
6 & 2.9:1 & 19 &  4.5 [0.2]\\% &5.70 [1.35] & 0.78 [0.19]\\%03/19 c
\hline
\hline
7 & 1.2:1 & 12 & 2.1 [0.1]\\% & & 0.82 [0.08] & 1.37 [0.18]\\%09/26 
8 & 1:1.0 &10 & 4.1 [0.1]\\% & 3.33 [0.37] & 1.22 [0.14]\\ %02/25
%8 & 1.5:1 & 10 & \nodata & 2.56 [0.40] & \nodata \\%03/11
10 & 1.1:1 & 10 & 3.6 [0.1]\\% & 2.80 [0.46] & 1.28 [0.21]\\%03/13
\hline
11 & 1:1.0 & 19 & 2.5 [0.2]\\% & 3.08 [0.62] & 0.80 [0.17]\\%03/21
\hline
12 & 1.1:1 & 24 & 2.7 [0.1]\\% & 2.53 [0.72] & 1.05 [0.30]\\%09/28 F
13 & 1:1.1 & 24 & 1.9 [0.1]\\% & 1.92 [0.77] &1.01 [0.41]\\%02/27
\hline
\hline
14 & 1:2.3 & 12 & 3.8 [0.1]\\% &  2.35 [0.40] & 1.60 [0.28]\\%08/20 F
15 & 1:3.1 & 10 & 3.2 [0.1]\\% & 0.96 [0.81] & 3.32 [2.79]\\%03/12 F
16 & 1:3.2 & 10 & 1.6 [0.1]\\% & 3.24 [0.75] & 0.50 [0.12]\\%03/26
\hline
17 & 1:3.2 & 20 & 3.9 [0.2]\\% & 1.06 [0.56] &3.69 [1.98]\\%08/21 F
18 & 1:2.8 & 19 & 4.1 [0.2]\\% & 2.89 [0.63] &1.42 [0.31]\\%03/20 F
\enddata
%\begin{list}{}
%\item $^1$If I use the fixed Gaussian for 210 min, k = 1.74 [0.09] %(the difference is negligible for conversion, though)
  %  \item $^2$If I use fixed Gaussians for 30-80 min and 210 min, k = 2.96 [0.13]
%    \item $^a$With python Gaussian fits instead of IDL numerical integration, 1.82 [0.15], ratio = 1.02 [0.10]  (this experiment has also a very low conversion compared to the rest). 
 %   \item $^c$With python Gaussian fits instead of IDL numerical integration, 1.21 [0.08], ratio = 1.33
%\end{list}{}
\tablecomments{Values in brackets correspond only to the fit errors. 
%, and the 20\% error for the column densities (see Sect. \ref{sec:exp-ir}) is not taken into account for 
%$N_{ss}$(NH$_2$CHO) does not affect the $k_{form}$ reaction rates since these are just a measure of how fast the steady state is reached, independently of the column density values. 
(The fit did not converge for Exp. 9.)}
\end{deluxetable}

\begin{deluxetable}{cccc} 
\tablecaption{Summary of the best-fit CH$_3$CHO formation cross-sections.\label{table_kineticsch3cho}}
\tablehead{
\colhead{Exp.} & \colhead{Ratio} & \colhead{Irr. T} & \colhead{\textit{$\sigma_{form}$}(CH$_3$CHO) }\\% & \colhead{$\sigma_{dest}$(CH$_4$)} & \colhead{$\sigma_{form}/\sigma_{dest}$}\\
\colhead{} & \colhead{} & \colhead{(K)} & \multicolumn{1}{c}{($\times$ 10$^{-18}$ cm$^{2}$)}
} 
%\colnumbers
\startdata
22 & 2.7:1 & 12 & 3.3 [0.2]\\% & 1.22 [0.24] & 2.71 [0.55] \\%05/31
23 & 2.1:1 & 12 & 5.2 [0.4]\\% & 1.06 [0.05] & 4.92 [0.41] \\%06/25
\hline
24 & 3.7:1 & 20 & 5.1 [0.8]\\% & 1.50 [0.25] & 3.42 [0.77] \\%06/26
\hline
\hline
25 & 1.2:1 & 12 & 4.4 [0.6]\\%& 1.09 [0.21] & 4.03 [0.94] \\%05/17
\hline
26 & 1:1.4 & 21 & 7.6 [0.7]\\% & 1.36 [0.31] & 5.58 [1.37] \\%05/30
\hline
\hline
27 & 1:2.6 & 11 & 2.2 [0.4]\\% & 1.36 [0.10] & 1.58 [0.29] \\%06/27
\hline
28 & 1:3.0 & 20 & 9.8 [0.6]\\% & 2.52 [0.55] & 3.87 [0.87] \\%08/06
\enddata
\tablecomments{Values in brackets correspond only to the fit errors.} 
%, and the 20\% error for the column densities (see Sect. \ref{sec:exp-ir}) is not taken into account for 
%$N_{ss}$(CH$_3$CHO) does not affect the $k_{form}$ reaction rates since these are just a measure of how fast the steady state is reached, independently of the column density values.}
\end{deluxetable}

\begin{figure}
    \centering
    \includegraphics[width=8cm]{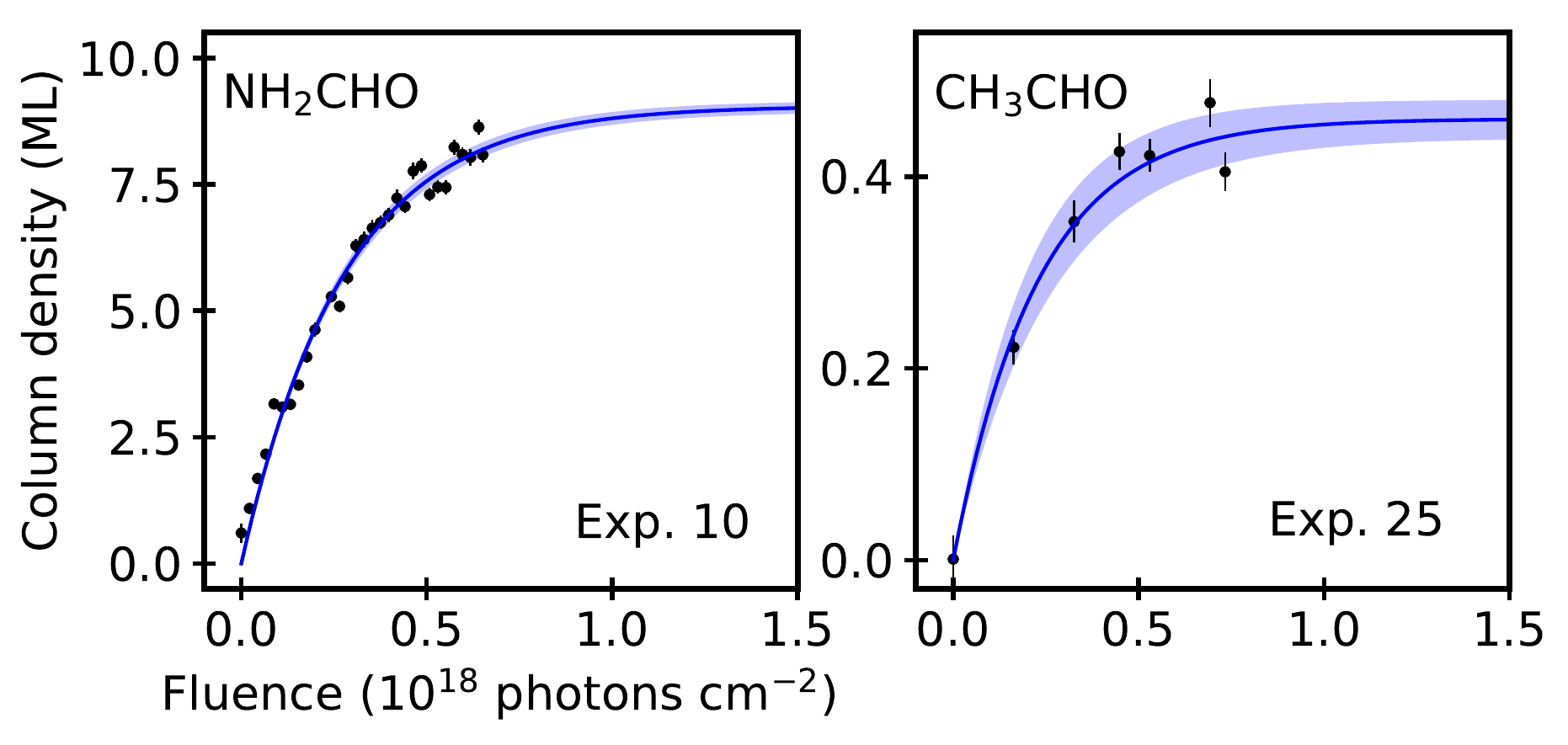}
    \caption{NH$_2$CHO and CH$_3$CHO growth curves (\textit{left} and \textit{right} panels, respectively) during photoprocessing of CO:NH$_3$ (Exp. 10) and CO:CH$_4$ (Exp. 25) ices (black dots), along with the best-fit kinetic model (blue) and the 1$\sigma$ confidence region for the fit (shadowed zone) (best-fit formation cross-sections are listed in Tables \ref{table_kineticsnh2cho} and \ref{table_kineticsch3cho}). 
Column densities are calculated from the spectra taken with an angle of 45$^{\circ}$ with respect to the infrared beam during irradiation of the ice samples (see Sect. \ref{sec:exp-ir}). \label{growth_fiducial}}
    
\end{figure}{}

\begin{figure}
    \centering
    \includegraphics[width=8cm]{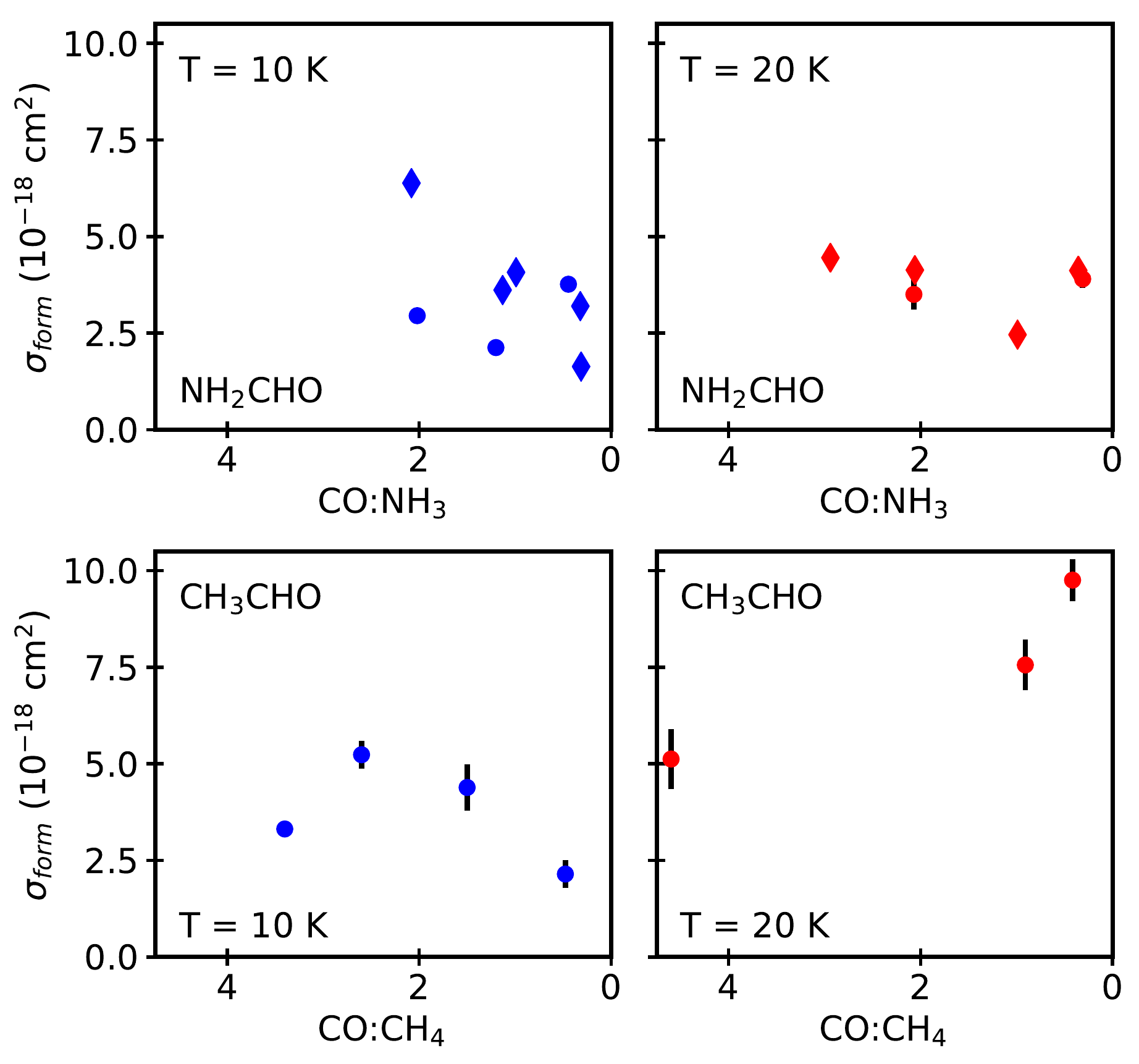}
    \caption{NH$_2$CHO and CH$_3$CHO formation cross-sections upon VUV irradiation of CO:NH$_3$ and CO:CH$_4$ ice mixtures (\textit{top} and \textit{bottom} panels, respectively). The cross-section values are listed in Tables \ref{table_kineticsnh2cho} and \ref{table_kineticsch3cho}, and represented by blue and red markers for the experimental simulations performed at T $\sim$ 10 K (\textit{left} panels) and $\sim$ 20 K \textit{right} panels), respectively; with error bars in black.
    We have used dots and diamonds for the first and the second series of CO:NH$_3$ experiments, respectively.\label{fig_sigmaform}}
 
\end{figure}

%The dissociation of NH$_3$ and CH$_4$ molecules leading to the formation of NH$_2^.$ and CH$_3^.$ radicals during energetic processing of the ice samples are unimolecular processes, and should thus follow a first-order kinetic equation \citep{jones11,bennett05}:

%\begin{equation}
%\begin{multiline}
%    N = (N_{0} - N_{ss})  \times e^{-\sigma_{dest}  Fluence} + N_{ss} 
 %   \end{equation}
   % or, alternatively
%    \begin{equation}
% N = (N_{0} - N_{ss}) \times e^{-k_{dest} t} + N_{ss},
%\end{multiline}
%\label{eqnh3}
%\end{equation}

%where $N_{0}$ is the initial ammonia or methane column density; $N_{ss}$ is their column density at the steady state; and $\sigma_{dest}$ is the apparent NH$_3$ or CH$_4$ destruction (or NH$_2^.$ and CH$_3^.$ formation) cross-section in cm$^2$. %; and $k_{dest}$ is the apparent rate constant in s$^{-1}$. 
%
The formation of NH$_2$CHO and CH$_3$CHO molecules upon processing of CO:NH$_3$ and CO:CH$_4$ ices can be parameterized with a pseudo-first order kinetic equation \citep{jones11,bennett05}: 

\begin{equation}
N = N_{ss} \times (1-e^{\sigma_{form} F})
%\end{equation}
%\begin{equation}
%N= N_{ss} \times (1-e^{k_{form} t}),
\label{eqnh2cho}
\end{equation}

where $N_{ss}$ represents the NH$_2$CHO or CH$_3$CHO steady state column densities, $\sigma_{form}$ is their apparent formation cross-section in cm$^2$, and $F$ the incident fluence in photons cm$^{-2}$. %; and $k_{form}$ the apparent formation rate in s$^{-1}$. %Since the kinetic equations are normalized to the column densities, $\sigma_{dest}$ and $\sigma_{form}$ are an indication of how fast the steady state is reached (i.e., how many photons are needed to reach the steady state), independently of the final column density values. 
Fig. \ref{growth_fiducial} presents the NH$_2$CHO and CH$_3$CHO growth curves (product column densities versus fluence) for the fiducial CO:NH$_3$ and CO:CH$_4$ 1:1 ice mixtures photoprocessed at 10 K (Exp. 10 and 25), along with the best-fit kinetic models. 
The NH$_2$CHO and CH$_3$CHO formation cross-sections are (3.6 $\pm$ 0.1) $\times$ 10$^{18}$ cm$^{-2}$ and (4.4 $\pm$ 0.6) $\times$ 10$^{18}$ cm$^{-2}$, respectively (see last columns of Tables \ref{table_kineticsnh2cho} and \ref{table_kineticsch3cho}). 
Within experimental uncertainties (note that experimental errors are larger than the fit errors), the formation kinetics of both COMs are similar under the same experimental conditions. Therefore, the higher NH$_2$CHO conversion yield at a given fluence (Sect. \ref{sec:conv}) is not due to a faster formation of NH$_2$CHO compared to CH$_3$CHO.

The NH$_2$CHO and CH$_3$CHO growth curves in experiments 2$-$18 and 22$-$28 are presented in 
%Figures \ref{growth_nh2cho} and \ref{growth_ch3cho}, respectively, 
the Appendix 
along with the best-fit kinetic models. 
The NH$_2$CHO and CH$_3$CHO formation cross-sections are listed in Tables \ref{table_kineticsnh2cho} and \ref{table_kineticsch3cho}, and shown in Fig. \ref{fig_sigmaform}. 
They range from 2 $\times$ 10$^{18}$ cm$^{-2}$ to 6 $\times$ 10$^{18}$ cm$^{-2}$, approximately, in the case of NH$_2$CHO, and 3 $\times$ 10$^{18}$ $-$ 10 $\times$ 10$^{18}$ cm$^{-2}$ for CH$_3$CHO, depending on the ice composition and irradiation temperature. The bottom line is that the formation kinetics of these COMs are always of the same order under all the experimental conditions explored in this work. 
We do not identify any consistent and significant trend for the COM formation cross-sections with the ice composition or irradiation temperature. 

We note that at 10 K we derive different cross sections by up to a factor of 3 in the two NH$_2$CHO formation experimental series. The inferred substantial experiment-to-experiment variation may mask small temperature or composition dependencies on the formation cross sections. We speculate that this is due to subtle fluctuations of the VUV lamp emission spectrum with time (Sect. \ref{sec:exp}), which demonstrates the limits inherent with studying photoprocesses with broadband lamps. It also points to the presence of non-thermal processes at 10 K that strongly depend on the specific photon energy. No difference across experimental series is seen for the 20 K experiments, or when considering formation efficiencies.

\section{Discussion}\label{discussion}

\subsection{COM formation mechanisms}\label{conversion}

In Sect. \ref{sec:conv} we show that the conversion yield from NH$_2^.$ to NH$_2$CHO is up to one order of magnitude higher than the conversion yield from CH$_3^.$ to CH$_3$CHO under similar experimental conditions. 
At the same time, the measured conversion yields follow opposite trends: 
%with the ice composition and irradiation temperature: 
while the NH$_2$CHO conversion yield increases with the NH$_3$ abundance and the irradiation temperature, the CH$_3$CHO conversion yield decreases with both the CH$_4$ abundance and the irradiation temperature. 
This suggests that even though the formation pathway is expected to be the same for both species \citep{jones11,bennett05}, their formation processes present some peculiarities that lead to the observed differences. 

%\subsubsection{Processes competing with the CH$_3$CHO formation}

%\begin{figure}
%\centering
%\includegraphics[width=6.5cm]{conversionratio.pdf}
%\caption{C$_{x}$H$_{2x+2}$ and CH$_3$CHO \textbf{column density ratio after a fluence of 6.5 $\times$ 10$^{17}$ photons cm$^{-2}$} at T $\sim$ 10 K (blue) and T $\sim$ 20 K (red) with error bars in black, .\label{fig_conversion_ethane}}
%When comparing column densities of two different species, the 20% error due to the band strength uncertainties should be taken into account. It is not included, though, in this figure since we focus on the conversion ratio changes from one composition to another, and not in the absolute value of the conversion ratio. 
%\end{figure}

\begin{figure}
    \centering
    \includegraphics[width=8cm]{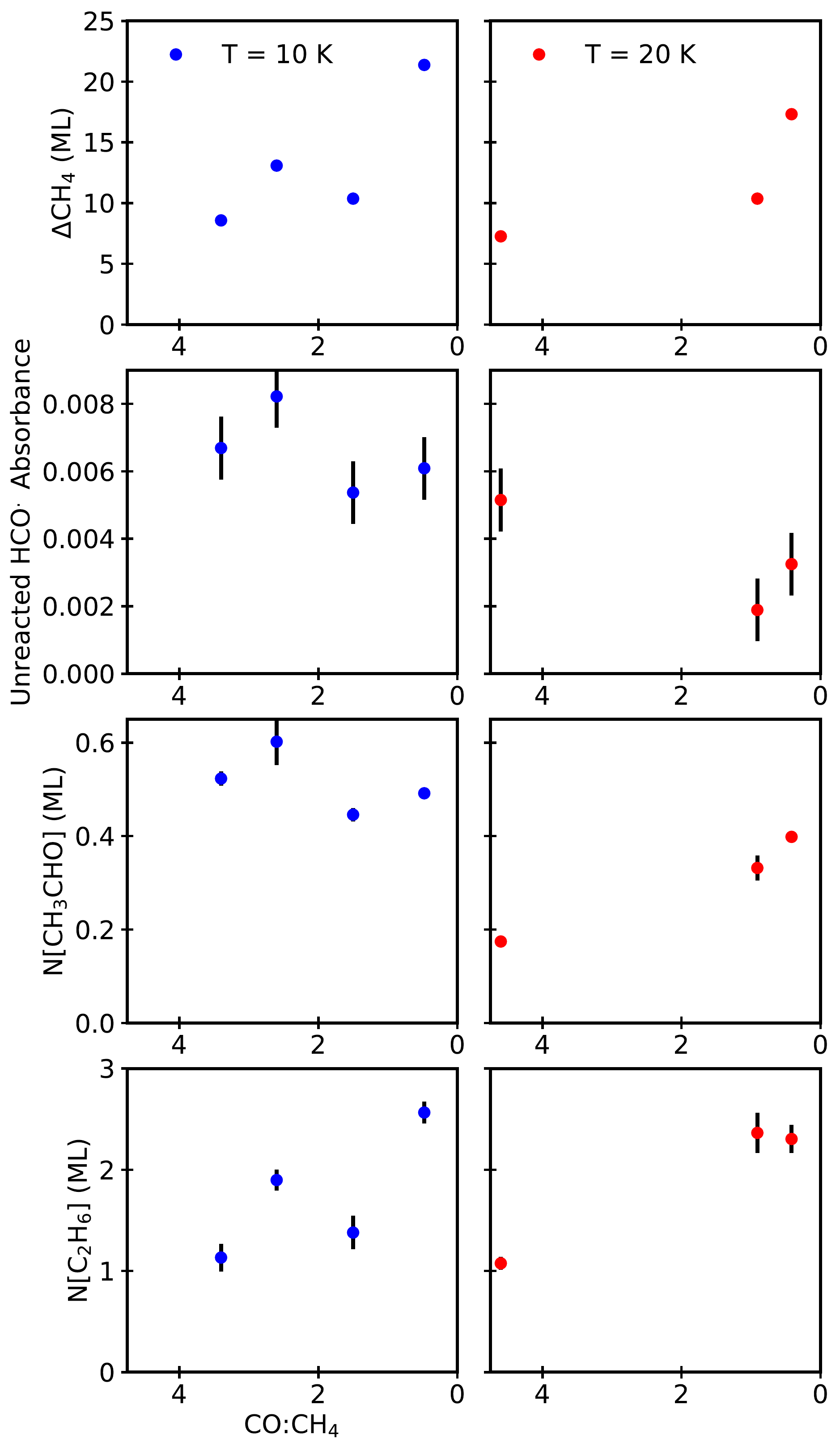}
    \caption{From top to bottom, number of photodestroyed CH$_4$ molecules; IR absorbance corresponding to the unreacted HCO$^.$ radicals; and column density of the photoproduced CH$_3$CHO and C$_2$H$_6$ \citep[based on the 1465 cm$^{-1}$ IR band, and using the band strength for CH$_4$-rich ices extracted from][]{german16}, after photoprocessing of CO:CH$_4$ ice samples at 10 K (left) and 20 K (right).}
    \label{fig:ch4}
\end{figure}{}

The main question that needs to be addressed is the much higher NH$_2$CHO conversion yield compared to the CH$_3$CHO yield. The NH$_2$CHO and CH$_3$CHO formation cross-sections are roughly similar (normalized to the steady state abundances, see Sect. \ref{sec:kin}) and cannot account for the order of magnitude difference in the conversion yields.   
%At the same time, while the formation of NH$_2$CHO reaches the steady state at similar time scales as the formation of NH$_2^.$ radicals during processing of CO:NH$_3$ ices ($k_{form}$/$k_{dest}$ $\sim$ 1, see top panels of Fig. \ref{fig_kratio}), the steady state for the formation of CH$_3$CHO is reached $\sim$2$-$6 times faster than the formation of CH$_3^.$ radicals (bottom panels of Fig. \ref{fig_kratio}). 
%We also note that HCO$^.$ radicals are detected in the IR spectrum of the photoprocessed CO:CH$_4$ ices, regardless the composition or the irradiation temperature (Sect. \ref{form_ch3cho}). This indicates that a fraction of these radicals remained unreacted in all experimental simulations with CO:CH$_4$ ice samples.  On the other hand, while HCO$^{.}$ radicals are also detected was not detected, though, in NH$_3$-rich ice samples, or after irradiation of any ice at T $\ge$ 20 K). 
%This suggests that 
A possible explanation is that there are 
more efficient processes that compete with the formation of CH$_3$CHO during the processing of CO:CH$_4$ ices, compared to the processes competing with the formation of NH$_2$CHO molecules in CO:NH$_3$ ices.  
%
%Since the CH$_3^.$ conversion to CH$_3$CHO decreases as the CH$_4$ abundance increases (bottom left panel of Fig. \ref{fig_conversion}), 
One of these competing processes is the recombination of CH$_3^.$ radicals to form larger saturated hydrocarbons\footnote{We note that simple hydrocarbons are usually detected in protostellar and circumstellar environments \citep[see, e.g.,][]{oberg08,pontoppidan14,guzman15}.} such as C$_2$H$_6$. These are indeed detected in the IR spectra of the photoprocessed CO:CH$_4$ ices (Fig. \ref{irch3cho}).
On the other hand, N$_2$H$_4$ is not detected after photoprocessing of the CO:NH$_3$ ice samples. This implies that the recombination of CH$_3^.$ radicals is probably more efficient than the recombination of NH$_2^.$ radicals, relatively to the formation of  CH$_3$CHO and NH$_2$CHO. %, \textbf{leading to overall higher conversion yields}.  

The efficient recombination of CH$_3^.$ radicals to form larger hydrocarbons (along with the inefficient recombination of NH$_2^.$ to form N$_2$H$_4$) could also explain some of the trends observed for the CH$_3$CHO and NH$_2$CHO conversion yields. %with the ice composition and the irradiation temperature, shown in Fig. \ref{fig_conversion}. 
Upon photoprocessing of CO:CH$_4$ ices of any composition at 10 K, a similar number of CH$_3$CHO molecules is produced in every experiment ($\sim$0.5 ML, Fig. \ref{fig:ch4}), while the number of produced CH$_3^.$ radicals increases with the CH$_4$ ice abundance  (see the increasing CH$_4$ photodisociation in the top left panel of Fig. \ref{fig:ch4}). 
%Figure \ref{fig:ch4} also shows the number of photodestroyed CH$_4$ molecules, that increases with the CH$_4$ abundance. However, this does not affect the number of CH$_3$CHO molecules formed. %causing the observed decrease of the CH$_3$CHO conversion yield at 10 K. 
The number of unreacted HCO$^.$ radicals in these experiments is roughly constant (Fig. \ref{fig:ch4}), so decreased HCO$^.$ production with increased CH$_4$ abundance cannot explain the decrease in the CH$_3$CHO conversion yield with CH$_4$ ice abundance.
On the other hand, the number of produced C$_2$H$_6$ molecules (based on the 1465 cm$^{-1}$ IR band shown in Fig. \ref{irch3cho}) follows the same trend observed for the number of photodestroyed CH$_4$ molecules, leading to a constant conversion yield of $\sim$13\%\footnote{Since it takes two CH$_3^.$ radicals to form one C$_2$H$_6$ molecule, this means that $\sim$26\% of the produced CH$_3^.$ radicals recombine to form C$_2$H$_6$ in any experiment.}, $\sim$2$-$5 times higher than the CH$_3$CHO conversion yield.
%of $\sim$13\% for all the experiments (i.e.,  
These trends can be understood if CH$_3^.$ radicals prefer to react with one another rather than with HCO$^.$ radicals when given the option, and considering that diffusion of radicals is not efficient at low temperatures in the absence of non-thermal, hot-atom diffusion, so only neighboring radicals can react at 10 K. 
Under these premises, CH$_3^.$ and HCO$^.$ radicals would need to form next to each other to form a CH$_3$CHO molecule, without any additional CH$_3^.$ radical that would result in the formation of C$_2$H$_6$. This is more likely to happen in CO-rich ices rather than in CH$_4$-rich ices, explaining the observed trend in the CH$_3$CHO conversion yield at 10 K. 
%In CH$_4$-rich ices the probability that two CH$_3^.$ radicals are produced in close proximity and subsequently recombine to form C$_2$H$_6$ is higher \textbf{than the probability that the CH$_3^.$ radical is formed close to a  HCO$^.$ radical. 
%This suggests that in the competition between hydrocarbon and CH$_3$CHO formation, CH$_3$CHO loses out unless the formed CH$_3^.$ is located next to a hydrogenated CO molecule. 
%An alternative explanation is that the number of photoproduced HCO$^.$ radicals decreases with the CH$_4$ abundance. However, in that case a decrease in the NH$_2$CHO conversion yield with the NH$_3$ abundance in the ice should be observed in Fig. \ref{fig_conversion} too, since the HCO$^.$ formation mechanism is expected to be the same in both ice samples (see Sect. \ref{sec:intro}). 

%
At higher temperatures, radicals are expected to become mobile \citep[see, e.g.,][]{garrod19}
and non-neighboring CH$_3^.$ radicals may also react with one another forming C$_2$H$_6$ molecules. The number of produced CH$_3$CHO molecules thus decrease down to $\sim$0.2 ML in CO-rich ices (see Table \ref{table_experiments_ch3cho} and Fig. \ref{fig:ch4}). This supports the hypothesis that CH$_3^.$ radicals preferentially react with one another instead that with HCO$^.$ radicals. However, the corresponding increase in the hydrocarbon production at 20 K compared to the production at 10 K is low, on the order of the C$_2$H$_6$ column density errors shown in the bottom panels of Fig. \ref{fig:ch4}.

\begin{figure*}
\centering\includegraphics[width=12.5cm]{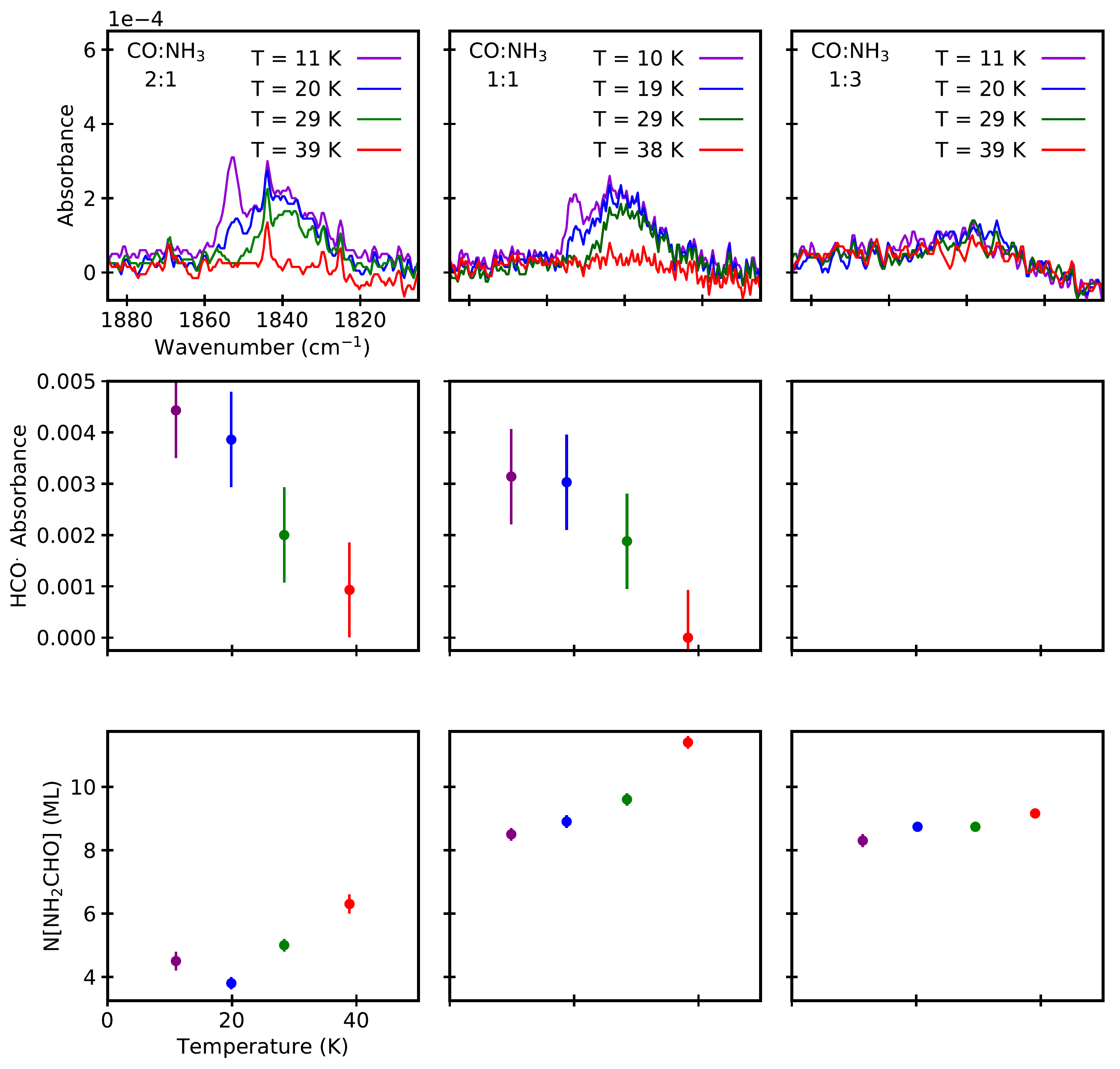}
\caption{The HCO$^.$ IR feature detected after photoprocessing of 2:1 (Exp. 2) and 1:1 (Exp. 10) CO:NH$_3$ ices at T = 10 K (\textit{left} and \textit{middle upper} panels) decreases its intensity at T $>$ 20 K (\textit{left} and \textit{middle middle} panels, the HCO$^.$ column density was not calculated to the lack of an experimental band strength value in the literature)}, due to the increased radical mobility that leads to an increase in the NH$_2$CHO column density (\textit{left} and \textit{middle bottom} panels). HCO$^.$ radicals are not detected in NH$_3$-rich ices (e.g., Exp. 14, \textit{right} panels).\label{ir_hco}
\end{figure*}

To explain the increase of the NH$_2$CHO conversion yield with the NH$_3$ ice abundance when the ice samples are irradiated at 10 K, we speculate that the HCO$^.$ and NH$_2^.$ radicals need a particular orientation for their reaction to proceed and produce NH$_2$CHO, and that this particular orientation is more likely to take place in NH$_3$-rich ices. 
Whether this is true also for the CH$_3^.$ + HCO$^.$ reaction is difficult to say with the present data, and we therefore focus on the NH$_2^.$ + HCO$^.$ reaction.
 
%In our experimental simulations, a
%HCO$^.$ radicals are not detected  upon photoprocessing of NH$_3$-rich ices as observed in the right top panel of Fig. \ref{ir_hco}. 
%That is, 
While some HCO$^.$ radicals remained unreacted after irradiation of the 2:1 and 1:1 CO:NH$_3$ ice samples at 10 K (left and middle upper panels of Fig. \ref{ir_hco}), no HCO$^.$ radicals were detected after irradiation of the NH$_3$-rich ice samples (right upper panel of Fig. \ref{ir_hco}). 
At the same time, an increase in the NH$_2$CHO conversion yield is observed. 
Therefore, all HCO$^.$ radicals formed in NH$_3$-rich ices are able to react with NH$_2^.$ radicals and form NH$_2$CHO.  
%The relative orientation of these radicals in water-rich ices is known to determine the final outcome of the NH$_2^.$ + HCO$^.$ reactivity (formation of NH$_2$CHO or reformation of NH$_3$ and CO), according to the theoretical quantum chemical computations presented in \citet{rimola18}. 
%
A possible explanation to this is that
the CO:NH$_3$ ratio decrease in the ice samples is accompanied by an increase of the NH$_2^.$:HCO$^.$ ratio during photoprocessing of the ice samples. In that case,  
%abundance of NH$_3$ molecules increases in the ice, 
a particular HCO$^.$ radical can be potentially surrounded by a larger number of NH$_2^.$ radicals. This would increase the probability that at least one of them is oriented in such way that enables the recombination of the two radicals and the formation of NH$_2$CHO\footnote{We note that, according to the reaction mechanism explained in Sect. \ref{sec:intro}, 
%for every HCO$^.$ radical formed in the ice samples, a NH$_2^.$ radical has been previously formed. In other words, 
it is not possible to form a larger number of HCO$^.$ than NH$_2^.$ radicals, while it is  possible to produce a larger number of NH$_2^.$ than HCO$^.$ radicals. Therefore, the opposite situation (i.e., a NH$_2^.$ radical surrounded by a large number of HCO$^.$ radicals) is much less probable in CO-rich ices.}. %, and thus the conversion yield. 
%Since for each HCO radical formed, a NH2 radical must be formed as well, it is not possible in the CO-rich ices to have a NH2 radical surrounded by a large number of HCO radicals, so this effect is not available in CO-rich ices. 
%%%%%%%%%%%% INCLUDE IF ASKED BY THE REFEREE %%%%%%%%%%%%%%%%%%%%%%%%%%%%%%%%%%%%%%%%%%%%%%%%%%%
%%For the same reason, a larger number of NH$_2^.$ radicals could remain unreacted in NH$_3$-rich ices, which would lower the value of the conversion yield. However, the recombination of H atoms and NH$_2^.$ radicals is more likely in NH$_3$-rich ices, which finally leads to higher conversion yields. %%SOMEHOW RELATED TO THE CAGE EFFECT 
%%%%%%%%%%%% INCLUDE IF ASKED BY THE REFEREE %%%%%%%%%%%%%%%%%%%%%%%%%%%%%%%%%%%%%%%%%%%%%%%%%%%
%
%
The HCO$^.$ IR feature detected after photoprocessing of 2:1 and 1:1 CO:NH$_3$ ices at 10 K decreases its intensity when the photoprocessed ice samples are warmed to temperatures above 20 K (left and middle upper panels of Fig. \ref{ir_hco}), and is not detected at T $\sim$ 40 K. 
At these temperatures, radicals become mobile, and the remaining HCO$^.$ could overcome this small reorientation barrier and find properly oriented NH$_2^.$ radicals. 
This leads to an increase of $\sim$30$-$40\% in the NH$_2$CHO column density (left and middle bottom panels inf Fig. \ref{ir_hco}), that increases the conversion yield to similar values as those found for NH$_3$-rich ices. %, and for ices of any composition irradiated at 20 K (
%%2:1 Increases from 4.3 to 6.3 (43%). An increase of 42% in the conversion of 19.2% leads to ~28%.
%%1:1 Increases from 8.5 to 11.4 (30%). An increase of 30% in the conversion of 24% leads to ~31%. 
%%This increase is not observed during warm-up of 1:3 CO:NH$_3$ ices, since no HCO$^.$ radicals remained unreacted after irradiation of NH$_3$-rich ices at T = 10 K. 
%
%\textbf{Even though no clear trend with the ice composition can be extracted for the NH$_2$CHO conversion yield in ices irradiated at 20 K (see Sect. \ref{sec:conv})}, 
We note that unreacted HCO$^.$ radicals are not detected when CO:NH$_3$ ice samples of any composition are irradiated at 20 K. 
%Therefore, no dependence on the ice composition should be expected for the NH$_2$CHO conversion yields at 20 K}. 
Therefore, NH$_2^.$ and HCO$^.$ radicals produced at 20 K must be mobile enough to react, regardless of the ice composition. This should result in a flat NH$_2$CHO conversion yield at 20 K. Unfortunately the conversion yield values measured for our 20 K experiments (upper left panel of Fig. \ref{fig_conversion}) had too much scatter to test this prediction, but we note that they appear to be compatible with a lack of dependence on NH$_3$ concentration.
At the same time, the large scatter in the estimation of the formation cross sections at different temperatures (top panels of Fig. \ref{fig_sigmaform}) prevents us from quantifying this small reorientation barrier. 

As stated above, the preferential recombination of CH$_3^.$ radicals to form larger hydrocarbons during photoprocessing
of CO:CH$_4$ ices does not allow us to check whether the radical orientation also plays a role in the formation of CH$_3$CHO molecules, but we note that  
the relative orientation of the HCO$^.$ and CH$_3^.$ radicals is supposed to be important for the production of CH$_3$CHO in water-rich ices, according to \citet{romero16}. %, that claimed that the reformation of CH$_4$ and CO would be the only possible outcome for the reaction of CH$_3^.$ and HCO$^.$ radicals, due to geometrical constraints imposed by the water ice surface. 
%\citet{romero16} claimed that acetaldehyde cannot be formed from CH$_3^.$ and HCO$^.$ radicals on icy grains due to the constraints on the radical orientation imposed by the water-ice matrix, according to theoretical quantum chemical simulations. However, \citet{lamberts19} proved that acetaldehyde can indeed be formed from CH$_3^.$ and HCO$^.$ on CO ice surfaces, depending on the initial orientation of the radicals.  

%
\subsection{Comparison with previous experiments}

A number of different NH$_2$CHO and CH$_3$CHO solid-state formation pathways have been previously explored experimentally.
In the laboratory, hydrogenation chemistry for the formation of NH$_2$CHO has had mixed results. 
Hydrogenation of HNCO was proven to be inefficient \citep{noble15}, 
but simultaneous hydrogenation of solid NO and H$_2$CO, and experiments that combine hydrogenation and UV irradiation of NO in a CO-rich ice analog result in small amounts of NH$_2$CHO production \citep{dulieu19,fedoseev16}. 

By contrast, it appears relatively easy to form both NH$_2$CHO and CH$_3$CHO upon photon or electron processing of different ices composed of common interstellar ice species. 
Warm-up of photoprocessed ice mixtures containing H$_2$O, CH$_3$OH, CO, and NH$_3$  \citep{bernstein95}, %NO INFORMATION ON THE FORMATION PATHWAY
UV and soft X-ray irradiation of a H$_2$O:CO:NH$_3$ ice analog \citep[respectively]{van04,ciaravella19}, 
UV and electron irradiation of ices containing H$_2$O, CH$_3$OH, and NH$_3$ at 5 and 70 K \citep{henderson15}, 
UV and electron irradiation of a binary CH$_3$OH:NH$_3$ ice mixture at 8$-$10 K \citep[respectively]{guillermo14,jheeta13}, 
electron irradiation of a CO:NH$_3$ ice mixture at 12 K \citep{jones11}, 
%electron irradiation of an H$_2$O:CH$_3$OH:CO$_2$:CH$_4$:NH$_3$ ice mixture at 20 K \citep{bergantini11}, 
%and ion irradiation of H$_2$O:CH$_4$:NH$_3$, H$_2$O:CH$_4$:N$_2$, and CH$_3$OH:N$_2$ ice mixtures at T$<$20 K \citep{kanuchova16},  %NO INFORMATION ON THE FORMATION PATHWAY 
all produce detectable amounts of NH$_2$CHO, among other products. 
%proton irradiation of H$_2$O:HCN and H$_2$O:NH$_3$:HCN ice mixtures at 18 K \citep{gerakines04}, 
%NO INFORMATION ON THE FORMATION PATHWAY
Formation of CH$_3$CHO is observed during 
UV irradiation of pure CH$_3$OH ices \citep{oberg09}  
and H$_2$O:CH$_4$ ice mixtures \citep{oberg10b}, 
and electron irradiation of a CO:CH$_4$ binary ice mixture at 10 K \citep{bennett05}.  

\subsubsection{Formation efficiencies upon UV processing of different ice samples}

%%%%% WARNING %%%%%%%%
%% THE CONVERSION YIELD DEPENDS ON THE FLUENCE, BUT IT DOES NOT NECESSARILY INCREASES WITH FLUENCE. THE BEHAVIOR DEPENDS ON THE NH3 AND CH4 DESTRUCTION CURVES, AND THE NH2CHO AND CH3CHO GROWTH CURVES. TO COMPARE DIFFERENT EXPERIMENTS IN THIS WORK, WE HAVE CHOSEN A 6.5E17 FLUENCE. TO COMPARE WITH OTHER WORKS, WE MUST COMPARE SIMILAR FLUENCES. 
%FOR THE SAME REASON, THE PERCENT YIELD ALSO DEPENDS ON THE FLUENCE
%I DON'T WANT TO GIVE CONVERSION YIELDS FOR DIFFERENT FLUENCES THAN 6.5E17 TO AVOID CONFUSION, AND I'M ONLY GIVING PERCENT YIELDS AT THOSE FLUENCES
%%%%% WARNING %%%%%%%%

We can compare the formation efficiencies at a particular UV fluence of those experiments mentioned above that report such values to those measured in our experiments at similar fluences, to obtain initial constraints on the relative formation efficiencies in different ice environments.   
%For the above experiments that quantify their formation yields, we can compare them with those  

After photoprocessing of the CO:CH$_4$ ice samples with a fluence of 3.2 $\times$ 10$^{17}$ photons cm$^{-2}$, the CH$_3$CHO percent yield  %ranges from 2\% to 5\% for the range of ice compositions and irradiation temperatures explored in this work (Sect. \ref{sec:conv}). 
(i.e., the number of CH$_3$CHO molecules formed with respect to the initial amount of parent CH$_4$ molecules) ranges from 0.2\% to 0.9\%, i.e., 0.2$-$0.9\% of the initial CH$_4$ is transformed into CH$_3$CHO for the range of ice compositions and irradiation temperatures explored in this work.  
This is on the same order of the CH$_3$OH and CH$_4$ transformed into CH$_3$CHO after $\sim$3 $\times$ 10$^{17}$ photons cm$^{-2}$ VUV irradiation of a pure CH$_3$OH ice and a H$_2$O:CH$_4$ ice mixture \citep[$\sim$0.5\%,][respectively]{oberg09,oberg10b}.
%%CH3CHO FORMATION CROSS SECTION IN OBERG 09 = <1E-18 (HERE IT IS HIGHER)
%% I DON'T THINK THE VALUE IS CALCULATED IN OBERG10B

On the other hand, the NH$_2$CHO conversion yield after a fluence of 6.5 $\times$ 10$^{17}$ photons cm$^{-2}$ ranges from 15\% to 37\% (Sect. \ref{sec:conv}), i.e., 3$-$8\% of the initial NH$_3$ is transformed into NH$_2$CHO in our experimental simulations.   
This is 
%compatible with the wide range of results found for other solid-state pathways: our NH$_2$CHO conversion yields in terms of the initial NH$_3$ are 
one order of magnitude higher than the value measured during warm-up of a photolyzed H$_2$O:CH$_3$OH:CO:NH$_3$ ice sample \citep[$<$1\%,][]{bernstein95} for a fluence 500 times higher.  
%%WARNING: KANUCHOVA COMPARES THE VALUE OF 1.8% OF THE SECOND ROW IN TABLE 2 OF BERNSTEIN ET AL. 1995, BUT I THINK WE SHOULD COMPARE THE FIRST ONE THAT ONLY INCLUDES FORMAMIDE AND ACETAMIDE
%but 2$-$5 times lower than the value reported in \citet{kanuchova16} for the ion irradiation of a H$_2$O:CH$_4$:NH$_3$ ice (up to 15\%). 
Assuming that the reaction mechanism is the same in both experiments (since both include CO and NH$_3$ molecules), and that both experiments are close to the steady state (see Fig. \ref{growth_nh2cho}), then the NH$_2^.$ to NH$_2$CHO conversion yield is decreased in the presence of a water-ice matrix. 
Therefore, the values presented in this work should be taken with caution, and this decrease should be taken into account when trying to extrapolate these results to the real astrophysical scenario.

\subsubsection{Comparison between UV- and electron-processed ices}\label{astroimp}

At the same time, we can compare the percent yields and formation rates upon VUV irradiation of CO:NH$_3$ and CO:CH$_4$ ice samples with those reported for the electron irradiation of the same ices in \citet{jones11} and \citet{bennett05}, since the depostied energies and energy fluxes are comparable. 

In \citet{jones11} and \citet{bennett05}, the energy deposited after 30$-$60 minutes of electron irradiation of the ice samples is 1.4$-$2.8 $\times$ 10$^{18}$ eV cm$^{-2}$. 
%also correspond to a timescale of $\sim$2 $\times$ 10$^{6}$ years. %, after scaling to the actual cosmic-ray energy deposition in dense clouds.  
%
In our experiments, a 3.2 $\times$ 10$^{17}$ photons cm$^{-2}$ photon fluence experienced by the ice samples correspond to a deposited energy of $\sim$2.5 $\times$ 10$^{18}$ eV cm$^{-2}$, assuming an average lamp photon energy of $\sim$7.75 eV ($\sim$160 nm, Fig. \ref{lampem}). %the secondary UV fluence experienced by interstellar ice mantles in dense clouds over a timescale of $\sim$2$-$4 $\times$ 10$^{6}$ years (Sect. \ref{sec:exp}). 
%Since the factory H$_2$D$_2$ lamp emission spectrum resembles the expected secondary UV field in the interior of dense clouds \citep{gredel89}, the energy deposited in our ice samples would also correspond to the energy deposited in interstellar ices over the same time span of $\sim$2$-$4 $\times$ 10$^{6}$ years. 
%
%Since the energy deposited by both cosmic rays and secondary UV photons in dense clouds over the same time span is similar (see above), the energy deposited in the CO:NH$_3$ ice samples during UV photoprocessing (this work) and electron irradiation \citep{jones11} is comparable. 
The NH$_2$CHO percent yield with respect to the initial amount of NH$_3$ measured at that fluence ranges from  2\% to  6\% under the different ice compositions and irradiation temperatures, which are comparable to the 1$-$2\% transformation after 2.8 $\times$ 10$^{18}$ eV cm$^{-2}$ dosed upon electron bombardment of CO:NH$_3$ ices in \citet{jones11}. The CH$_3$CHO percent yield with respect to the initial amount of CH$_4$ was not reported, though, in \citet{bennett05}. 

The 1.4$-$2.8 $\times$ 10$^{18}$ eV cm$^{-2}$ deposited energy on a time span of 30$-$60 minutes correspond to an energy flux of 7.7 $\times$ 10$^{14}$ eV cm$^{-2}$ s$^{-1}$ during the electron irradiation of the ice samples in \citet{jones11} and \citet{bennett05}. The reported NH$_2$CHO and CH$_3$CHO formation rates are $\sim$5 $\times$ 10$^{-4}$ s$^{-1}$, which result in formation cross-sections of $\sim$6.5 $\times$ 10$^{-19}$ cm$^{-2}$ per incident eV. 
The formation cross-sections presented in Tables \ref{table_kineticsnh2cho} and \ref{table_kineticsch3cho} range from 2 to 5 $\times$ 10$^{-18}$ cm$^{-2}$ per incident photon under most of the experimental conditions explored in this work, which correspond to 2.6$-$6.5 $\times$ 10$^{-19}$ cm$^{-2}$ per incident eV, assuming the average lamp photon energy of $\sim$7.75 eV, i.e., on the same order than the formation cross-sections found upon electron bombardment of the same ice samples. 
%In order to calculate the formation rates from the formation cross-sections presented in Sect. \ref{sec:kin}, we multiply by the VUV flux of our H$_2$D$_2$ lamp (Sect. \ref{sec:exp}). % and the equation \ref{equvflux}:
%
%\begin{equation}
%    k_{form} = \sigma_{form} \times UV Flux
%    \label{equvflux}
%\end{equation}{}
%
%The measured NH$_2$CHO formation rates upon VUV photoprocessing of CO:NH$_3$ ices range from 0.6 $\times$ 10$^{-4}$ s$^{-1}$ to 3.1 $\times$ 10$^{-4}$ s$^{-1}$, while the CH$_3$CHO formation rates range from 1.3 $\times$ 10$^{-4}$ s$^{-1}$ to 5.6 $\times$ 10$^{-4}$ s$^{-1}$,
%on the same order of magnitude than the formation rates found upon electron bombardment of the same ice samples \citep[$\sim$5 $\times$ 10$^{-4}$ s$^{-1}$,][]{jones11,bennett05}. 
%Similar reaction rates for the formation of different species during ion bombardment and UV photolysis of H$_2$O-rich ices had already been reported \citep[see, e.g.,][]{moore01}. 
%Previous experimental simulations have proven that different types of energetic processing of ice samples simulating photon or cosmic ray irradiation of interstellar ices lead to similar end products, although some differences are found \citep{islam14,guillermo14}. 

\subsection{Astrophysical implications}

In dense clouds, cosmic rays induce both a secondary UV field by their interaction with the gas-phase H$_2$ molecules \citep{cecchi92,shen04}, and secondary keV electrons when interacting with ice molecules \citep{hovington97}. 
The energy dose experienced by the ice mantles from the cosmic-ray induced secondary UV field is thought to be similar in value to the energy contribution from the incoming cosmic rays \citep{moore01}. 
Therefore, since the ice formation efficiencies and timescales of NH$_2$CHO and CH$_3$CHO upon both kinds of energetic processing are similar (see above), the contribution of the photoproduced NH$_2$CHO and CH$_3$CHO in interstellar ices should to be comparable to the cosmic-ray induced production.

%\subsection{Comparison with observed abundances}

As stated in the introduction, the observed NH$_2$CHO-to-CH$_3$CHO abundance ratio may help us constrain the origin of these species. 
In our experimental simulations, the conversion yield from NH$_2^.$ to NH$_2$CHO is up to one order of magnitude higher than the conversion yield from CH$_3^.$ to CH$_3$CHO (Sect. \ref{sec:conv}) upon VUV photoprocessing of CO:NH$_3$ and CO:CH$_4$ ice samples, respectively. Formation of CH$_3$CHO in ice mantles seems to be quenched by the competing formation of larger hydrocarbons, compared to the formation of NH$_2$CHO molecules. 
We expect a similar NH$_2$CHO:CH$_3$CHO ratio for the production of these COMs in interstellar ice mantles, since the results found in this work should be applicable to all ices where NH$_2^.$, CH$_3^.$, and HCO$^.$ radicals were generated (Sect. \ref{sec:intro}), and the observed ice abundances of NH$_3$ and CH$_4$ with respect to solid H$_2$O are similar \citep{boogert15}. 
If we assume that interstellar ices are completely desorbed in hot corinos around low-mass protostars, so that the difference in the NH$_2$CHO and CH$_3$CHO desorption temperatures does not affect the subsequent gas-phase abundances, we would then expect an order of magnitude higher abundances for NH$_2$CHO compared to CH$_3$CHO in such regions, opposite to what is observed  \citep[$\sim$0.1 NH$_2$CHO/CH$_3$CHO abundance ratio,][]{lee19}. This suggests that there are additional CH$_3$CHO formation pathways contributing to the CH$_3$CHO formation in addition to CH$_3^.$ + HCO$^.$ chemistry in both the water-rich ice mantles (e.g., CH$_3$OH ice processing), and in the gas-phase. NH$_2$CHO may on the other hand originate from energetic ice processing; more observational constraints are needed to differentiate between this and competing pathways.

\section{Conclusions}\label{conclussions}

\begin{enumerate}

    \item NH$_2$CHO and CH$_3$CHO are formed upon VUV photoprocessing of CO:NH$_3$ and CO:CH$_4$ ice mixtures, respectively.  The conversion from radicals to COMs is 2$-$16 times higher for NH$_2$CHO compared to CH$_3$CHO. This is likely due to the competing formation of saturated hydrocarbons that prevents the reaction of CH$_3^.$ and HCO$^.$ radicals to form CH$_3$CHO. 
    
    \item %A higher radical mobility at T $\ge$ 20 K compared to 10 K leads to an increase in the conversion yield from NH$_2^.$ to NH$_2$CHO. The conversion yield also increases with the NH$_3$ abundance in the ice. 
    The observed trends in the NH$_2$CHO conversion yield with the ice composition and the irradiation temperature are compatible with the proposal in \citet{romero16} that a particular orientation of the radicals (NH$_2^.$ and HCO$^.$ in this case) is needed for the formation of NH$_2$CHO, with a small reorientation barrier that can be overcome with the temperature. 
    The preferential formation of larger hydrocarbons prevents us from determining whether there is a similar barrier for the formation of CH$_3$CHO. %, and a decreased conversion yield for CH$_3^.$ to CH$_3$CHO. %%, while in the CO:CH$_4$ samples the competing formation of saturated hydrocarbons is further favored. 
    %This wouldn't be the case for the CH$_3$CHO formation, but cannot be confirmed since the formation of larger hydrocarbons is favored in both cases. 
    %\item The production of CH$_3$CHO takes place in time scales 2$-$3 times shorter than the production of NH$_2$CHO. In any case, t
    
    \item  The formation efficiencies and rates for NH$_2$CHO and CH$_3$CHO upon VUV photoprocessing of CO:NH$_3$ and CO:CH$_4$ ice samples are similar to those found upon electron bombardment of the same ice mixtures for similar deposited energies. We thus expect a comparable contribution from UV photons and cosmic rays for the formation of these species through the radical combination pathways in interstellar ices. 
    
    \item Given the measured relative conversion yields for NH$_2$CHO and CH$_3$CHO, a significant contribution of other ice- or gas-phase CH$_3$CHO formation pathways are needed to explain the observed NH$_2$CHO/CH$_3$CHO ratio in hot corinos ($\sim$0.1).
    
\end{enumerate}

\acknowledgments

This work was supported by an award from the Simons Foundation (SCOL \# 321183, KO).

\begin{figure}{}
\centering
\includegraphics[width=8cm]{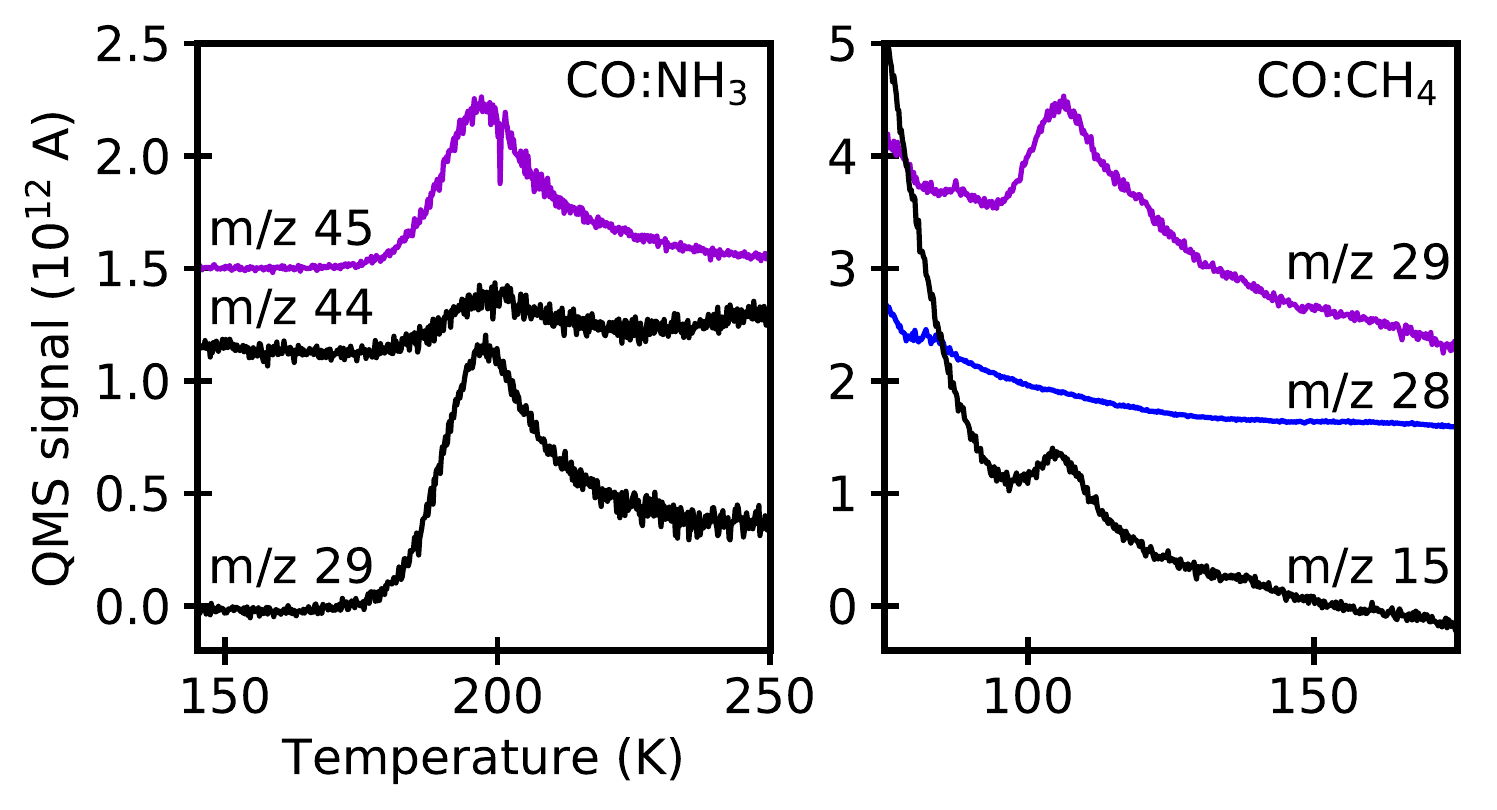}
\caption{TPD curves of the main mass fragments (violet) corresponding to the produced NH$_2$CHO (\textit{left panel}) and CH$_3$CHO (\textit{right panel}), and the following most intense fragments (black) after photoprocessing of a CO:NH$_3$ and a CO:CH$_4$ ice sample, respectively (see Figures \ref{tpdnh2cho} and \ref{tpdch3cho}). The $m/z$ = 28 mass fragment corresponding to CO is also shown in blue in the right panel. \label{tpd_app}}
\end{figure}

\appendix
Fig. \ref{tpd_app} shows the TPD curves of the main mass fragments corresponding to NH$_2$CHO (left panel) and CH$_3$CHO (right panel) after photoprocessing of a CO:NH$_3$ and a CO:CH$_4$ ice sample, respectively. 

\begin{figure*}{}
\centering
\includegraphics[width=18cm]{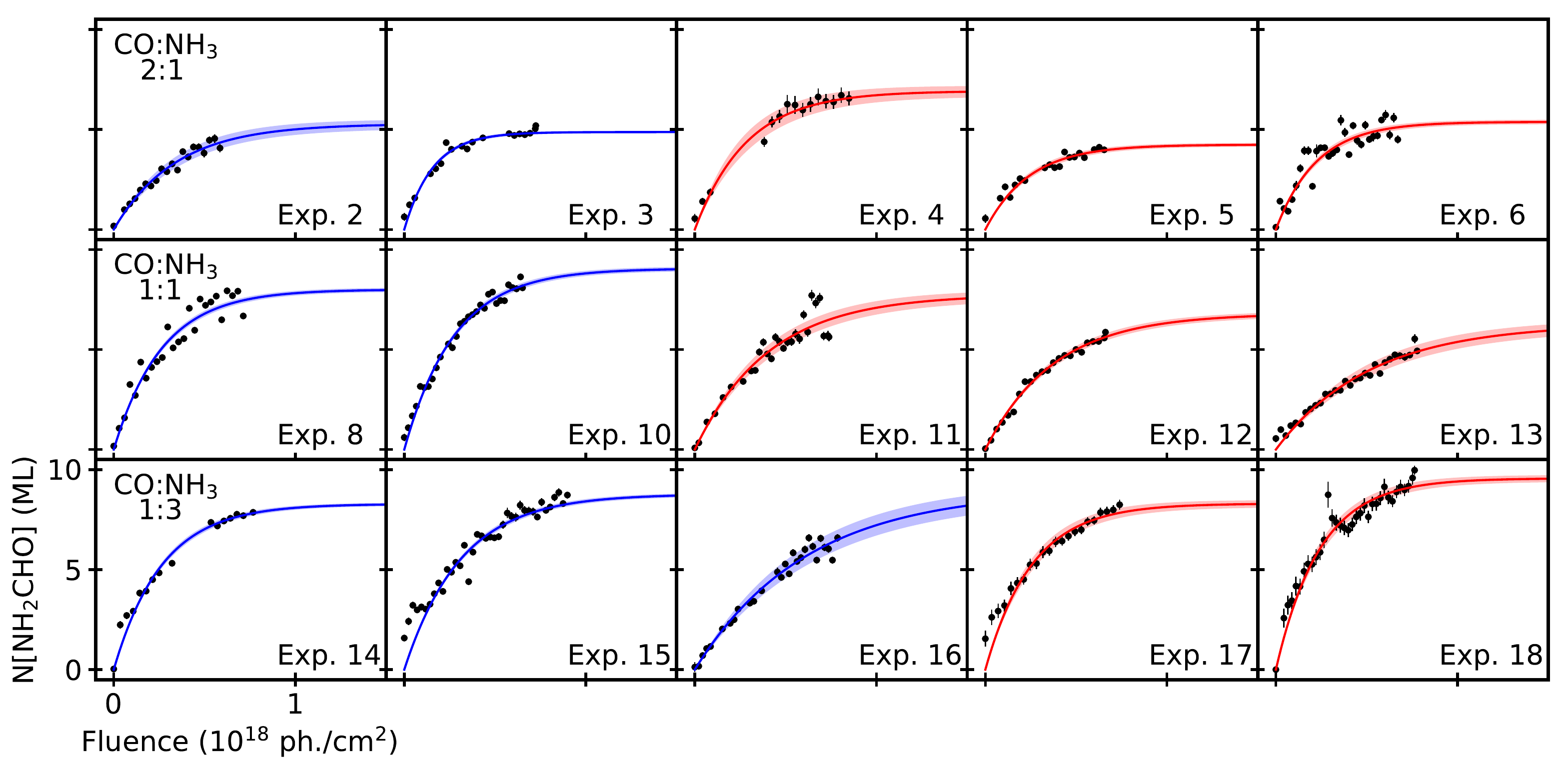}
\caption{NH$_2$CHO growth curves during photprocessing of CO:NH$_3$ ices in Exp. 2$-$18 (black dots), along with the best-fit pseudo-first order kinetic model (blue and red curves for the $\sim$ 10 K and $\ge$ 20 K ice photoprocessing experiments, respectively) and the 1$\sigma$ confidence region for the fit (shadowed zone). The best-fit parameters and 1$\sigma$ fit errors are listed in Table \ref{table_kineticsnh2cho}. 
Column densities are calculated from the spectra taken with an angle of 45$^{\circ}$ with respect to the infrared beam during photoprocessing of the ice samples (see Sect. \ref{sec:exp-ir}).
Exp. 7 and 9 are not shown for an easier visualization of the experiments performed with similar ice compositions but different irradiation temperatures.\label{growth_nh2cho}}
\end{figure*}

\begin{figure}{}
\centering
\includegraphics[width=8cm]{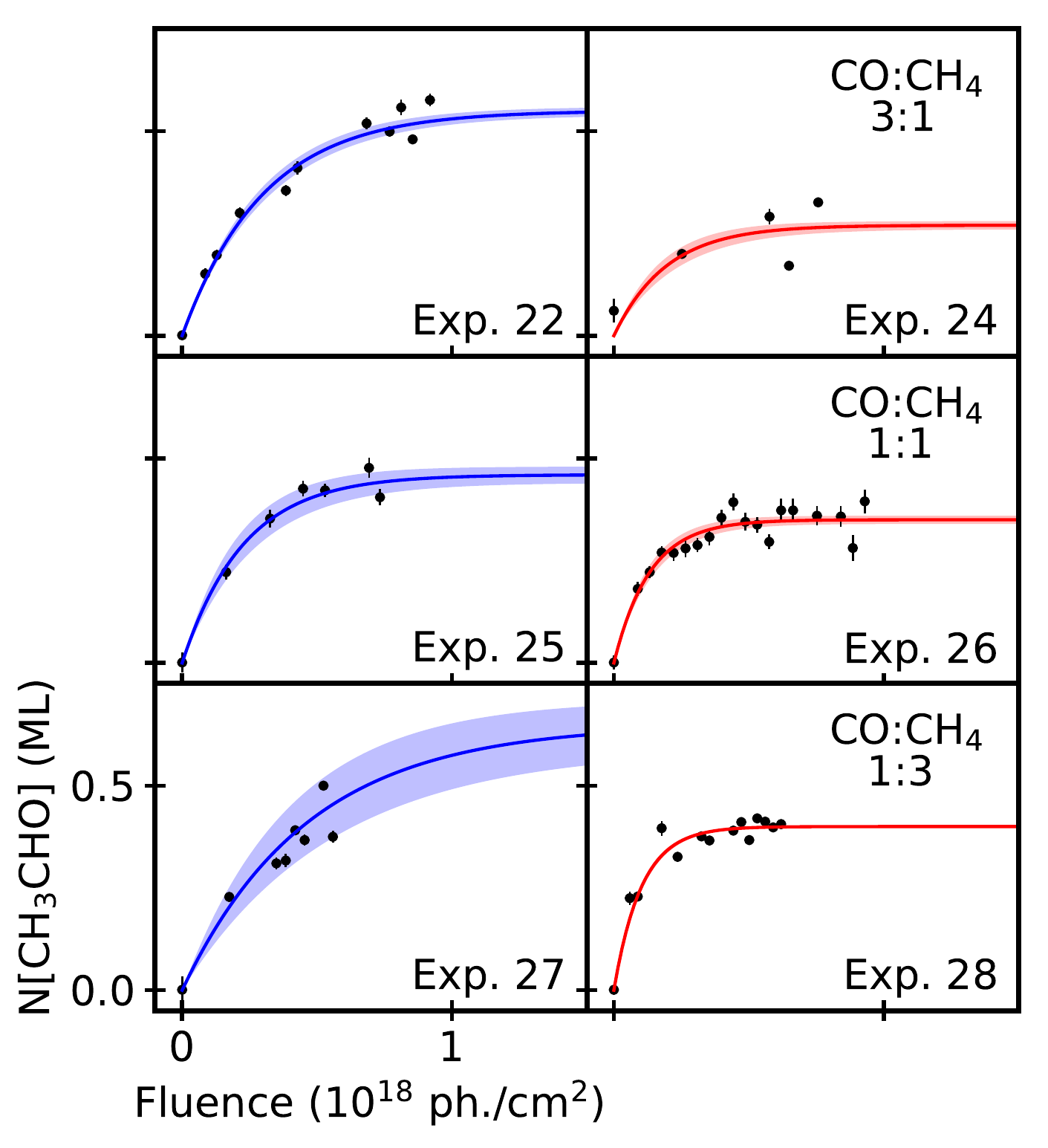}
\caption{CH$_3$CHO growth curves during photprocessing of CO:CH$_4$ ices in Exp. 22$-$28 (black dots), along with the best-fit first order kinetic model (blue and red curves for the $\sim$ 10 K and $\sim$ 20 K ice photoprocessing experiments, respectively) and the 1$\sigma$ confidence region for the fit (shadowed zone). The best-fit formation cross-sections and 1$\sigma$ fit errors are listed in Table \ref{table_kineticsch3cho}. 
Column densities are calculated from the spectra taken with an angle of 45$^{\circ}$ with respect to the infrared beam during photoprocessing of the ice samples (see Sect. \ref{sec:exp-ir}). 
Exp. 22 is not shown for an easier visualization of the experiments performed with similar ice compositions but different irradiation temperatures.\label{growth_ch3cho}}
\end{figure}

The NH$_2$CHO and CH$_3$CHO growth curves for the experiments in Tables \ref{table_experiments_nh2cho} and \ref{table_experiments_ch3cho} are presented in Figures \ref{growth_nh2cho}, and \ref{growth_ch3cho}.  

%\begin{figure*}{}
%\centering
%\includegraphics[width=15cm]{kinetic_fit_all_nh3.pdf}
%\caption{Ammonia destruction curves during photoprocessing of CO:NH$_3$ ices in Exp. 2$-$17 (black dots), along with the best-fit first order kinetic model (blue and red curves for the $\sim$ 10 K and $\ge$ 20 K ice photoprocessing experiments, respectively) and the 1$\sigma$ confidence region for the fit (shadowed zone). The best-fit apparent destruction rate constants and 1$\sigma$ fit errors are listed in Table \ref{table_kineticsnh2cho}. 
%Column densities are calculated from the spectra taken with an angle of 45$^{\circ}$ with respect to the infrared beam during irradiation of the ice samples. 
%We note that the large uncertainties in Experiments 14 and 16 are caused by the poorly constrained steady state column density due to the very slow destruction rates found in those cases.}
%\label{growth_nh3}
%\end{figure*}

%\begin{figure}{}
%\centering
%\includegraphics[width=7cm]{kinetic_fit_all_ch3cho.pdf}
%\caption{}
%\label{growth_ch3cho}
%\end{figure}



\end{document}